\documentclass[journal]{IEEEtran}
\IEEEoverridecommandlockouts
\usepackage{cite}
\usepackage{amsmath,amssymb,amsfonts}
\usepackage{algorithm}
\usepackage{algpseudocode}
\usepackage{graphicx}
\usepackage{psfrag}

\graphicspath{{./images/}}
\DeclareGraphicsExtensions{.pdf,.jpeg,.png}
\usepackage{textcomp}
\usepackage{xcolor}
\def\BibTeX{{\rm B\kern-.05em{\sc i\kern-.025em b}\kern-.08em
    T\kern-.1667em\lower.7ex\hbox{E}\kern-.125emX}}
\newcommand{\tildebf}[1]{\Tilde{\mathbf{#1}}}

\newcounter{tempEquationCounter}
\newcounter{thisEquationNumber}

\begin{document}

\title{Low-Complexity Reliability-Based Equalization and Detection for OTFS-NOMA
}

\author{\IEEEauthorblockN{Stephen McWade, \textit{Member, IEEE,}, Arman Farhang, \textit{Senior Member, IEEE,} and
Mark F. Flanagan, \textit{Senior Member, IEEE}}
\thanks{S. McWade was with the School of Electrical and Electronic Engineering, University College Dublin, Belfield, Dublin 4, D04 V1W8 Ireland. He is now with the Department of Electronic and Electrical Engineering, Trinity College Dublin, Dublin 2, D02 PN40 Ireland (email: smcwade@tcd.ie). A. Farhang is with the Department of Electronic and Electrical Engineering, Trinity College Dublin, Dublin 2, D02 PN40 Ireland (email: arman.farhang@tcd.ie). M. F. Flanagan is with the School of Electrical and Electronic Engineering, University College Dublin, Belfield, Dublin 4, D04 V1W8 Ireland (email:mark.flanagan@ieee.org).

This publication has emanated from research conducted with the financial support of Science Foundation Ireland (SFI) under Grant number 17/RC-PhD/3479, Grant number 17/US/3445 and Grant number 19/FFP/7005(T).
}}

\maketitle

\begin{abstract}
    Orthogonal time frequency space (OTFS) modulation has recently emerged as a potential 6G candidate waveform which provides improved performance in high-mobility scenarios. In this paper we investigate the combination of OTFS with non-orthogonal multiple access (NOMA). Existing equalization and detection methods for OTFS-NOMA, such as minimum-mean-squared error with successive interference cancellation (MMSE-SIC), suffer from poor performance. Additionally, existing iterative methods for single-user OTFS based on low-complexity iterative least-squares solvers are not directly applicable to the NOMA scenario due to the presence of multi-user interference (MUI). Motivated by this, in this paper we propose a low-complexity method for equalization and detection for OTFS-NOMA. The proposed method uses a novel reliability zone (RZ) detection scheme which estimates the reliable symbols of the users and then uses interference cancellation to remove MUI. The thresholds for the RZ detector are optimized in a greedy manner to further improve detection performance. In order to optimize these thresholds, we modify the least squares with QR-factorization (LSQR) algorithm used for channel equalization to compute the the post-equalization mean-squared error (MSE), and track the evolution of this MSE throughout the iterative detection process. Numerical results demonstrate the superiority of the proposed equalization and detection technique to the existing MMSE-SIC benchmark in terms of symbol error rate (SER).    
\end{abstract}


\section{Introduction}
The sixth generation (6G) of mobile networks is expected to support communications in high-mobility environments such as high-speed rail, vehicle-to-everything (V2X) and unmanned aerial vehicle (UAV) communications \cite{Tataria_6G}. Orthogonal frequency division multiplexing (OFDM) has been the waveform utilized in the 4th and 5th generation of wireless networks. However, it is well-known that in high-mobility scenarios, OFDM performs poorly due to the Doppler effect \cite{Wei_OTFS}. In recent years, a new waveform called orthogonal time frequency space (OTFS) has been proposed to address this drawback of OFDM in time-varying channels. In contrast to OFDM, which transmits data symbols in the time-frequency domain, OTFS places the data symbols in the delay-Doppler domain \cite{Hadani_OTFS}. OTFS then uses a transformation to spread each information symbol over the whole time-frequency plane. This means that the symbols are all equally affected by the time and frequency selectivity of the channel which converts the time-varying channel to a time-invariant one in the delay-Doppler domain.

 A number of OTFS equalization and detection schemes have been proposed in the literature in recent years. The majority of these methods can be categorized into either low-complexity linear equalizers \cite{Tiwar_OTFS_LMMSE,Surabhi_OTFS_EQ,ZOU_OTFS_EQ} or non-linear message-passing-based equalizers \cite{Rav_INT_canc_MP, Surabhi_MP_mmWAVE, MIMO_OTFS_MP}. However, such methods assume a scattering environment in which the channel impulse response is sparse in the delay-Doppler domain. Under more realistic channel conditions, the low-complexity linear schemes are no longer applicable as the assumptions they make about the channel no longer hold. Additionally, message-passing-based detectors become prohibitively complex due to the large number of scatterers \cite{Qu_OTFS_detection}. An alternative approach was proposed in \cite{Qu_OTFS_detection} which utilized a least-squares minimum residual (LSMR) based
channel equalizer and a reliability-based
dynamic detector. However, the system model in \cite{Qu_OTFS_detection} only considers a single-user scenario and it is not applicable to the multi-user scenario that is of interest in this paper.

For a multi-user OTFS system, the multiple access (MA) technique utilized is an important consideration. How best to multiplex users in the delay-Doppler domain is an open question and there have been numerous recent works which propose different methods \cite{patent_Hadani,surabhi2019multiple,Chong_OTFS_Uplink_Rate}. These methods can be broadly categorized into orthogonal multiple access (OMA)  or non-orthogonal multiple access (NOMA). In OTFS-OMA, users are multiplexed either in the delay domain or the Doppler domain, and only one user can occupy a given resource block \cite{Chong_OTFS_Uplink_Rate}. However, the users suffer from multi-user interference (MUI) due to the Doppler spread, which degrades performance. MUI can be mitigated by inserting guard bands between users, as was done in \cite{patent_Hadani}. However, this use of guard bands leads to a spectral efficiency (SE) loss \cite{surabhi2019multiple}. 

An alternative approach is OTFS-NOMA, where the users are allowed to occupy the same resource block and are multiplexed in either the power domain or the code domain. A multi-user detection (MUD) scheme, such as successive interference cancellation (SIC), is then used to detect the user symbols \cite{DaiSurveyNOMA}. NOMA is a well-known technique which can provide improved SE over the corresponding OMA system as well as potentially higher connectivity as the number of users supported by a NOMA system is not limited by the number of physical resources available. A number of OTFS-NOMA schemes have been proposed in the literature in recent years that use either power-domain \cite{Poor_OTFS_NOMA, Chatt_OTFS_NOMA} or code-domain \cite{OTFS_SCMA, OTFS_code_noma} multiplexing. This paper focuses on power-domain OTFS-NOMA. 

With regard to the existing work on power-domain OTFS-NOMA, the authors of \cite{Poor_OTFS_NOMA} considered a single high-mobility OTFS user multiplexed with multiple low-mobility OFDM users. However, this system model is restricted to a single OTFS user and hence cannot accommodate multiple high-mobility users. The authors of \cite{Chatt_OTFS_NOMA} addressed this issue and proposed an OTFS-NOMA scheme which utilizes a rectangular pulse shape where multiple users overlap in the delay-Doppler domain and are multiplexed in the power domain. The results presented in \cite{Chatt_OTFS_NOMA} show that OTFS-NOMA achieves higher spectral efficiency than the equivalent OTFS-OMA system. However, the system proposed in \cite{Chatt_OTFS_NOMA} used minimum-mean-squared-error (MMSE) equalization in combination with SIC for equalization and detection. The problem with this scheme is that direct implementation of MMSE equalization is prohibitively computationally complex and thus impractical for real-world scenarios. 

As of yet, to the best of our knowledge, there is no low-complexity equalization and detection method for power-domain OTFS-NOMA. In addition, the low-complexity equalization and detection method of \cite{Qu_OTFS_detection} for single-user OTFS is not directly applicable to a NOMA scenario due to the presence of MUI. This paper addresses these gaps in the literature with the following contributions: 
\begin{itemize}
    \item We propose a novel iterative method for equalization and detection of a downlink OTFS-NOMA system which, within each iteration, uses a proposed modified LSQR (mLSQR) algorithm to equalize the channel, an RZ detector to detect reliable symbols from both users, and interference cancellation to improve detection on subsequent iterations.
    \item Our proposed modified LSQR algorithm, in addition to equalizing the channel, also computes the post-equalization MSE of the users' symbols, in contrast to the conventional LSQR algorithm. We derive an exact closed-form expression for this MSE as well as a low-complexity approximation which capitalizes on the properties of the delay-Doppler channel in OTFS systems.
    \item We use a novel, greedy approach for optimizing the RZ thresholds within each iteration. This is in contrast to other RZ schemes which use heuristic thresholds \cite{Qu_OTFS_detection, Taubock_LSQR_2011}. Our method works by tracking the post-equalization MSE after interference cancellation and optimizing the RZ thresholds in each iteration to minimize the MSE. 
\end{itemize}

Additionally, we present numerical results which compare the SER performance of the proposed equalization and detection method with the existing MMSE-SIC benchmark \cite{Chatt_OTFS_NOMA}. We also compare the performance of our optimized RZ threshold design to a pre-determined threshold design. The presented results demonstrate the superiority of our proposed method, especially for the NOMA user with the smaller power allocation. A preliminary version of this work was described in \cite{mcwade_OTFS}, which showed the advantage of this general approach but did not include the derivation of the post-equalization MSE (or its low-complexity approximation), and also did not show how this MSE could be utilized to optimize the RZ thresholds in each iteration.

 The rest of this paper is organized as follows. Section~II describes the system model for a 2-user OTFS-NOMA system. In Section~III, the proposed equalization and detection algorithm is presented. Section~IV  describes the modified LSQR algorithm which equalizes the channel and computes the post-equalization MSE. Section~V presents the process for optimizing the thresholds of the RZ detector. Section~VI presents numerical results. Finally, Section~VII concludes the paper. 

 \subsubsection*{Notations} Superscripts ${(\cdot)^{\rm{T}}}$ and ${(\cdot)^{\rm{H}}}$ denote transpose and Hermitian transpose, respectively. Bold lower-case characters are used to denote vectors and bold upper-case characters are used to denote matrices. The function $\rm{vec}\{\mathbf{X}\}$ vectorizes the matrix $\mathbf{X}$ by stacking its columns to form a single column vector, and $\otimes$ represents the Kronecker product. The $p\times{p}$ identity matrix and $p \times q$ all-zero matrix are  denoted by $\mathbf{I}_p$ and $\mathbf{0}_{p\times{q}}$, respectively.
 
\section{System Model}
For ease of exposition, in the following sections we will describe the system model and the proposed detector for the case of a 2-user downlink OTFS-NOMA system; however, note that with appropriate modifications, the proposed method is applicable to any number of users. We consider a downlink OTFS-NOMA system where both users occupy the same delay-Doppler domain resources and are multiplexed in the power domain. For User $i \in \{ 1,2 \}$, let the $M \times N$ matrix $\mathbf{X}_i$ contain the $MN$ quadrature amplitude modulation (QAM) data symbols placed in the delay-Doppler domain. The elements of $\mathbf{X}_i$ are assumed to be independent and identically distributed (i.i.d.) complex random variables. Additionally, a normalized (unit-energy) square QAM constellation is assumed for each user.

In the first stage of OTFS modulation, the inverse symplectic fast Fourier transform (ISFFT) is used to map the delay-Doppler data symbols in $\mathbf{X}_i$ to the time-frequency domain. The ISFFT can be implemented by performing an $M$-point DFT operation on each of the columns of $\mathbf{X}_i$ followed by an $N$-point IDFT operation on each of the rows of $\mathbf{X}_i$. The time-frequency signal matrix of User $i$ is therefore given by
\begin{equation}
    \mathbf{D}_i = \mathbf{F}_{M}\mathbf{X}_i\mathbf{F}_{N}^{\rm{H}},  \label{eq:1}
\end{equation}
where $\mathbf{F}_N$ is the $N$-point unitary discrete Fourier transform (DFT) matrix in which the $(l,k)$ element is $\frac{1}{\sqrt{N}}e^{-j\frac{2\pi}{N}lk}$. Next, cyclic prefix OFDM (CP-OFDM) modulation is used to convert the time-frequency signal to the delay-time domain. The OTFS transmit signal matrix is therefore given by
\begin{equation}
    \mathbf{S}_i = {\mathbf{A}_{\mathrm{cp}}}\mathbf{F}_{M}^{\rm{H}}\mathbf{D}_i, \label{eq:2}
\end{equation}
 where ${\mathbf{A}_{\mathrm{cp}}} = \left[\mathbf{J}_{\rm{cp}}, \mathbf{I}_{N}\right]$ is the CP addition matrix (here $\mathbf{J}_{\rm{cp}}$ is composed of the last $N_{\mathrm{cp}}$ rows of $\mathbf{I}_{N}$). Using (\ref{eq:1}), the delay-time domain transmit signal can be rewritten as
\begin{equation}
    \mathbf{S}_i = {\mathbf{A}_{\mathrm{cp}}}\mathbf{X}_i\mathbf{F}_{N}^{\rm{H}}, \label{eq:3}
\end{equation}
and thus OFDM-based OTFS reduces to an $N$-point IDFT operation on the rows of $\mathbf{X}_i$ \cite{Arman_OTFS_2018}. 
After parallel to serial conversion, the time-domain symbols for User $i$ can now be written as
\begin{equation}
    \mathbf{s}_i = \mathrm{vec}(\mathbf{S}_i) =  (\mathbf{F}_{N}^{\mathrm{H}}\otimes{\mathbf{A}_{\mathrm{cp}}})\mathbf{x}_i.\label{eq:4}
\end{equation}
The users are multiplexed in the power domain and their signals are superimposed before transmission. The superimposed transmit signal is given by
\begin{equation}
    \mathbf{s} = \sqrt{\rho_1}\mathbf{s}_1 + \sqrt{\rho_2}\mathbf{s}_2,
    \label{eq:5}
\end{equation}
where $\rho_i$ is the power allocation coefficient for  for User i, and $\rho_1 + \rho_2  = 1$ (these power allocation coefficients are determined using an appropriate power allocation scheme, such as that used in [16]). We consider user indices to be ordered in descending order of their power allocation coefficients, i.e., $\rho_1 > \rho_2$.

After digital to analog conversion, the continuous-time signal $s(t)$ is transmitted through the linear time-varying (LTV) channel. The received signal at the receiver for User $i \in \{ 1,2 \}$ can be written as
\begin{equation}
        r_i(t) = \int \int h_i(\tau,\nu)s(t-\tau)e^{j2\pi\nu(t-\tau)}d\tau d\nu + \omega_i(t)\label{eq:6}
\end{equation}
where $$h_i(\tau,\nu) = \sum_{p=0}^{P_i-1}h_{i,p}\delta(\tau - \tau_{i,p})\delta(\nu - \nu_{i,p}),$$ is the delay-Doppler channel impulse response (CIR) for User $i$, which consists of $P_i$ channel paths, and $\omega_i(t)$ is the complex AWGN with variance $\sigma_{i}^2$. The parameters $h_{i,p}$, $\tau_{i,p}$ and $\nu_{i,p}$ represent the channel gain, delay and Doppler shift, respectively, associated with path $p$ of User $i$'s channel. The power delay profile (PDP) of the channel of User $i$ is given by $\boldsymbol{\lambda}_i = [\lambda_i(0), \ \dots  \ , \lambda_i(0)]$ and is assumed to be normalized such that $\sum_{p=0}^{P_i-1}\lambda_i(p) = 1$. Each channel path gain is modeled as a complex Gaussian random variable with mean zero and and variance $\lambda_i(p)$. Since the PDP is considered to be normalized, the average received SNR of User $i$ is given by $\mathrm{SNR}_i = \frac{p_1 + p_2}{\sigma_i^2}.$.  We assume perfect knowledge of the User $i$ channel at the receiver of User $i$, as previously considered in \cite{Chatt_OTFS_NOMA}. 

The received signal is then sampled with sampling period $T_{\rm{s}}$ and the discrete received signal samples can be expressed as 
 \begin{equation}
        r_i[n] =  \sum_{l=0}^{L-1}h_i[n,l]s[n-l] + \omega_i[n], \label{eq:7}
\end{equation}
where $h[n,l]$ is the CIR at time instant $n$ and delay $l$. The discrete-time received signal can be written in matrix form as 
\begin{equation}
        \mathbf{r}_i = \mathbf{H}_i\mathbf{s} + \boldsymbol{\omega}_i, \label{eq:8}
\end{equation}
where $\boldsymbol{\omega}_i$ is the complex AWGN vector and $\mathbf{H}_i$ is the $MN \times MN$ time-domain channel matrix of User $i$ constructed from the CIRs. The received signal is then demodulated and converted back to the delay-Doppler domain by taking an $N$-point DFT operation across the time domain samples. Thus, the received signal is given by 
\begin{equation}
        \mathbf{y}_i = (\mathbf{F}_{N}\otimes\mathbf{R}_{\mathrm{cp}})\mathbf{r}_i \label{eq:9}.
\end{equation}
This can alternatively be written as 
\begin{equation}
        \mathbf{y}_i = \mathbf{G}_i\mathbf{x}_{\mathrm{sup}} + \mathbf{w}_i \label{eq:10}.
    \end{equation}
where $\mathbf{G}_i=(\mathbf{F}_{N}\otimes\mathbf{R}_{\mathrm{cp}})\mathbf{H}_i(\mathbf{F}_{N}^{\mathrm{H}}\otimes\mathbf{A}_{\mathrm{cp}})$ is the effective channel matrix, $\mathbf{x}_{\mathrm{sup}}=\sqrt{\rho_1}\mathbf{x}_1 + \sqrt{\rho_2}\mathbf{x}_2$ is the superimposed delay-Doppler symbol vector and $\mathbf{w}_i=(\mathbf{F}_{N}\otimes\mathbf{R}_{\mathrm{cp}})\boldsymbol{\omega}_i$ is the noise vector. 


\section{Proposed Equalization and Detection Technique}

Each user needs to equalize the channel and detect its own symbols at its own receiver. One way to do this is to use MMSE equalization in combination with SIC, as in \cite{Chatt_OTFS_NOMA}, which we refer to as MMSE-SIC. MMSE equalization operates by pre-multiplying the received vector $\mathbf{y}_i$ in (\ref{eq:10}) by the MMSE equalization matrix given by
\begin{equation}
    \mathbf{W}_{\mathrm{MMSE},i} = \left((\mathbf{G}_i^{\mathrm{H}}\mathbf{G}_i+\sigma_{i}^2\mathbf{I})^{-1}\right)\mathbf{G}_i^{\mathrm{H}}. \label{eq:11}
\end{equation}
More specifically, User 1 uses $\mathbf{W}_{\mathrm{MMSE},1}$ to equalize the channel and then detect its own data symbols while treating the User 2 data symbols as noise. On the other hand, User 2 uses $\mathbf{W}_{\mathrm{MMSE},2}$ to equalize the channel and first detect the User 1 symbols, treating its own symbols as noise. User 2 then removes the User 1 signal from the received signal, uses $\mathbf{W}_{\mathrm{MMSE},2}$ to equalize the channel and then detects its own data symbols \cite{Chatt_OTFS_NOMA}. MMSE equalization is impractical for real-world applications due to the $MN \times MN$ matrix inversion in (\ref{eq:11}), which has a computational complexity of $\mathcal{O}(M^3N^3)$. This is clearly unrealistic for practical applications where $M$ and $N$ can be large. Additionally,  while low-complexity implementations of MMSE equalization exist, they assume ideal pulses and a small number of channel scatterers, and thus are not applicable to practical scenarios \cite{Qu_OTFS_detection}. 

The proposed method is inspired by the method proposed in \cite{Qu_OTFS_detection} for single-user OTFS which utilized an iterative LSMR-based method with RZ detection and interference cancellation. Note that if the method in \cite{Qu_OTFS_detection} is applied directly to OTFS-NOMA with SIC to detect the signals of User 1 and User 2, we can expect poor performance due to the MUI present in the system. Therefore, in the proposed method, we perform SIC at a symbol level rather than a packet level as is done in the MMSE-SIC approach. This allows for the decoding of symbols from both users as soon as they become reliable and also allows for the incorporation of MUI cancellation to improve the detection performance. The proposed algorithm uses an iterative process in which  the mLSQR algorithm is used to equalize the channel and an RZ detector is used to detect the reliable symbols of both User 1  \textit{and} User 2 within each iteration. Interference cancellation is then used to remove ISI, IDI \textit{and} MUI from the undetected symbols of both users, which improves the detection quality in subsequent iterations. The proposed mLSQR algorithm, which equalizes the channel and also computes the post-equalization MSE, will be explained in detail in Section~IV. In the next subsection, we describe the RZ detection process.

\subsection{Reliability zone detector}

\begin{figure}[t]
    \centering
    \includegraphics[width = 0.95\columnwidth]{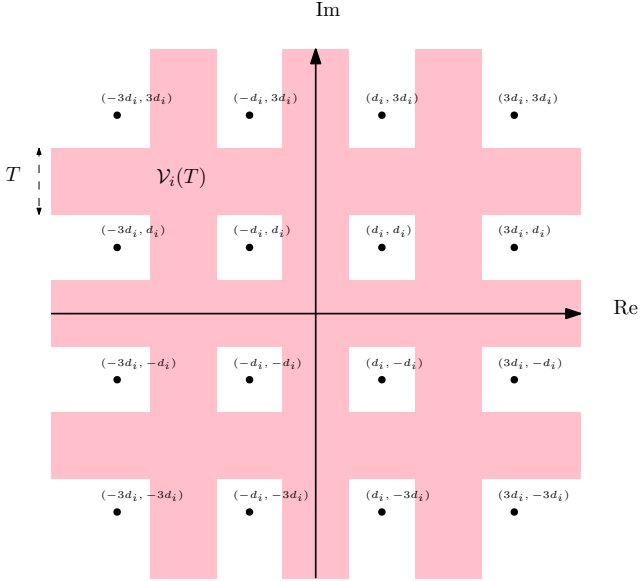}
    \caption{Illustration of the unreliable zone $\mathcal{V}_i(T)$ in  the case where User i employs a 16-QAM constellation.}
    \label{fig:RZ}
\end{figure}

Here, we first introduce some relevant notation. Each User $i \in \{1,2 \}$ uses $A_i$-ary QAM modulation, where the QAM symbol constellation is defined as 
\begin{equation*}
    \begin{split}
        \mathcal{A}_i =  \{&u + vj \ : \  u,v \in \{(2a-1)d_i : 
        \\& a \in \{-\sqrt{A_i}/2 +1 , \ \dots \ , \sqrt{A_i}/2 \}\}\},
    \end{split}
\end{equation*}
where $d_i$ is half the distance between adjacent QAM constellation symbols of User $i$ (the value of $d_i$ is chosen so as to ensure a unit-energy constellation $\mathcal{A}_i$). Next, we define the \textit{unreliable zone} with respect to this QAM constellation as
\begin{equation}
    \mathcal{V}_i(T) =  \{u + vj \ | u,v \in \mathcal{U}_i(T) \}, \label{eq:12}
\end{equation}
where
\begin{equation}
    \mathcal{U}_i(T)  =  \bigcup_{a=-\sqrt{A_i}/2 +1}^{\sqrt{A_i}/2 -1} \mathcal{U}_{i,a}(T) .
\end{equation}
and $$\mathcal{U}_{i,a}(T)  = \{u \ | \ 2ad_i-T/2 < u < 2ad_i + T/2 \},$$ where $T$ is a pre-defined threshold which determines the size of the unreliable zone. To demonstrate, the shaded areas in Fig. \ref{fig:RZ} shows an illustration of the unreliable zone $\mathcal{V}_i(T)$ for a 16-QAM constellation.

In the detection process, decisions are made in a symbol-by-symbol manner. If a symbol $x_i[n]$ is outside $\mathcal{V}_i(T_i)$, then it is deemed reliable and can be quantized to the nearest symbol in $\mathcal{A}_i$; the resulting symbol is denoted by $x_{i,q}[n] = Q_i(x[n])$. If $x_i[n]$ is inside $\mathcal{V}_i(T_i)$ then it is deemed unreliable and no quantization takes place. The detected reliable symbols can then be used for interference cancellation.

\begin{algorithm}[t]
\caption{Proposed Algorithm for symbol detection at User~$i$ receiver}\label{PA1}
\begin{algorithmic}[1]
\State \textbf{Input}: User index $i$, Channel matrix $\mathbf{G}_i$, received symbol vector $\mathbf{y}_i$, power allocation fractions $\rho_1$ and $\rho_2$
\State \textbf{Initialize}: $\mathbf{y}^{(1)}=\mathbf{y}_i$, $\mathbf{\hat{x}}_1= \mathbf{\hat{x}}_2 = \tildebf{x}_{1,\mathrm{q}}= \tildebf{x}_{2,\mathrm{q}}= \mathbf{0}_{MN\times1}$ 
\State Define $\mathcal{N}=\{0,\dots,MN-1\}$, $\mathcal{N}_1=\mathcal{N}_2=\mathcal{N}$, $\mathcal{D}_1= \emptyset$
\For{$k$ = 1 to $K$}
\State $[\tildebf{x}_{\mathrm{sup}}, \gamma]=\mathrm{mLSQR}(\mathbf{G}_i, \mathbf{y}^{(k)}, \sigma_i^2)$

\State $\tildebf{x}_{1} = ((\tildebf{x}_{\mathrm{sup}}[m])_{m\in\mathcal{N}_1})/\sqrt{\rho_1}$
\State $\tildebf{x}_{2}=((\tildebf{x}_{\mathrm{sup}}[m])_{m\in \mathcal{N}_2 \cap  \mathcal{D}_1})/\sqrt{\rho_2}$


\State \textbf{if} $i=1$
\State \ \ \ \  Select threshold $T_1$ by solving (\ref{eq40})
\State \ \ \ \  Select threshold $T_2$ by solving (44)
\State \textbf{else if} $i=2$
\State \ \ \ \  Select threshold $T_1$ by solving (48)
\State \ \ \ \  Select threshold $T_2$ by solving (\ref{eq40})
\State \textbf{end if}

\State Update users' reliable symbol index sets via $$\mathcal{R}_j = \{n \in \mathcal{N}_j :  \tildebf{x}_{j}[n] \notin \mathcal{V}_j(T_j^{(k)}) \}, \forall j \in \{1,2\}$$

\State Quantize reliable symbols: $$\tildebf{x}_{j,\mathrm{q}}[r] = Q_j(\tildebf{x}_{j}[r]), \ \forall r \in \mathcal{R}_j, \forall j \in \{1,2\}$$

\State Remove interference:
$$\mathbf{y}^{(k+1)}=\mathbf{y}^{(k)}-\mathbf{G}_i(\sqrt{\rho_1}\tildebf{x}_{1,\mathrm{q}} + \sqrt{\rho_2}\tildebf{x}_{2,\mathrm{q}})$$

\State Store detected symbols in output vectors:
$$\mathbf{\hat{x}}_j = \tildebf{x}_{j,\mathrm{q}}[r], \ \forall r \in \mathcal{R}_j, \forall j \in \{1,2\}$$

\State Reset: $\tildebf{x}_{1,\mathrm{q}}=\mathbf{0}$ and $\tildebf{x}_{2,\mathrm{q}}=\mathbf{0}$

\State Update: 
$\mathcal{N}_1=\{n \in \mathcal{N}: {\hat{x}}_1[n]=0\}$, $\mathcal{N}_2=\{n \in \mathcal{N}: {\hat{x}}_2[n]=0\}$, $\mathcal{D}_1=\{n \in \mathcal{N}: n \notin \mathcal{N}_1\}$

\State \textbf{if} $\mathcal{N}_i=\emptyset$, \textbf{break}

\EndFor
\State \textbf{Output}: $\mathbf{\hat{x}}_i$ 

\end{algorithmic}
\end{algorithm}

\subsection{Proposed algorithm}
In this subsection, we describe the proposed method for equalization and detection of the OTFS-NOMA signal at the receiver of User $i \in \{ 1,2 \}$. This method is described in Algorithm \ref{PA1}. Each iteration begins on line 5 of Algorithm 1, where the LSQR algorithm is used to equalize the channel and obtain a new estimate, $\tildebf{x}_{\mathrm{sup}}$, of the superimposed transmitted symbol vector via 
\begin{equation}
    [\tildebf{x}_{\mathrm{sup}}, \gamma_i]=\mathrm{mLSQR}(\mathbf{G}_i, \mathbf{y}^{(k)}, \sigma_i^2).
\end{equation}
Additionally, our proposed modification to the LSQR algorithm calculates the post-equalization MSE (denoted by $\gamma$) over all of the symbols of both users. The exact workings of the mLSQR algorithm and the role of the MSE $\gamma$ in optimizing the RZ detector's thresholds will be explained in detail in Sections IV and V, respectively. In lines 6 and 7, two sub-vectors are formed from $\tildebf{x}_{\mathrm{sup}}$. 
The vector $\tildebf{x}_{1}$ contains the elements of $\tildebf{x}_{\mathrm{sup}}$ whose indices are in $\mathcal{N}_1$, which is the set of undetected User 1 symbols. Since the RZ detector can only make decisions on User 2 symbols once the corresponding User 1 symbols have been detected on a previous iteration, the vector $\tildebf{x}_{2}$ contains the elements of $\tildebf{x}_{\mathrm{sup}}$ whose indices are in $\mathcal{N}_2$, the set of undetected User 2 symbols, \textit{and} $\mathcal{D}_1$, the set of detected User 1 symbols. In lines 8 -- 14, Algorithm 1 selects the thresholds, $T_1$ and $T_2$, to be used in the RZ detector. The exact process for selecting the thresholds will be explained in detail in Section~V. Decisions are then made on the reliability of the estimated symbols in $\tildebf{x}_{1}$ and $\tildebf{x}_{2}$ via the RZ detector in line 15.

In line 16, the reliable symbols are quantized to the nearest QAM symbol and are stored in the empty vectors $\tildebf{x}_{1,\mathrm{q}}$ and $\tildebf{x}_{2,\mathrm{q}}$. In line 17, the quantized reliable symbols are used to remove interference from the received signal vector via
\begin{equation}
    \mathbf{y}^{(k+1)}=\mathbf{y}^{(k)}-\mathbf{G}_i(\sqrt{\rho_1}\tildebf{x}_{1,\mathrm{q}} + \sqrt{\rho_2}\tildebf{x}_{2,\mathrm{q}}).
\end{equation}
The quantized symbols are also stored in the estimated symbol vectors $\hat{\mathbf{x}}_1$ and $\hat{\mathbf{x}}_2$ (line 18). After canceling the interference from the detected symbols of both users, the algorithm updates the sets  $\mathcal{N}_1$ and $\mathcal{N}_2$ of undetected symbols, and the set  $\mathcal{D}_1$ of detected User 1 symbols, based on the state of the output vectors $\mathbf{\hat{x}}_1$ and $\mathbf{\hat{x}}_2$. Since this is at the User $i$ receiver, the algorithm stops when all of the User $i$ symbols are detected, i.e., User 1 will detect all of its symbols before it detects all the User 2 symbols and can therefore stop once $\mathbf{\hat{x}}_1$ has no entries equal to zero.

Clearly, the performance of the RZ detector and the interference cancellation depend heavily on the thresholds $T_1$ and $T_2$. To the best of the authors' knowledge, in all existing works in the literature which use RZ detection, the thresholds are pre-determined and are reduced geometrically in each iteration \cite{Qu_OTFS_detection, Hampeis_LSQR, Taubock_LSQR_2011}. However, in the NOMA context the performance of a user can be significantly affected by the MUI from the other user (especially for the user with lower power allocation). Hence, it is beneficial to optimize the thresholds $T_1$ and $T_2$ to improve the detection performance. Consequently, we use a greedy approach in which $T_1$ and $T_2$ are optimized within each iteration; for this, the post-equalization MSE, $\gamma$, is needed. The conventional LSQR algorithm of \cite{LSQR} does not provide this, and therefore a modified LSQR algorithm is proposed in the following section.

In this paper we focus on the 2-user case as this allows for greater simplicity and clarity in our analysis. However, while Algorithm 1 is presented for the case of 2 users, it can be modified in a straightforward manner to deal with the case of $J$ users where $J \ge 2$, as follows. First, the sets $\mathcal{N}_j=\{n \in \mathcal{N}: {\hat{x}}_j[n]=0\}$ and $\mathcal{D}_j = \mathcal{N} \backslash \mathcal{N}_j$ are defined for each User $j \in \mathcal{J}$, where $ \mathcal{J} = \{1,2,\ldots,J\}$. Second, Line 7 in Algorithm 2 is replaced by a loop which sets $\tildebf{x}_{j}=((\tildebf{x}_\mathrm{sup}[m])_{m\in \mathcal{N}_j \cap  \mathcal{D}_{j-1}})/\sqrt{\rho_j}$ for each $j=2$ to $J$. Finally, Lines 8-14 in Algorithm 1 are replaced by a loop where, for each $j \in\mathcal{J}$, threshold $T_j$ is determined by solving (40) if $j=i$, by solving (44) if $j>i$, and by solving (48) if $j<i$. Here, the references to (44) and (48) refer to these optimization problems with appropriately modified user indices. 

\section{Modified LSQR Algorithm}

In this section, we present our proposed modified version of the LSQR algorithm, which is listed in Algorithm 2. We begin by summarizing the basic operation of the conventional LSQR algorithm, which remains unchanged in Algorithm 2. Then we describe the proposed modification which computes the post-equalization MSE. Two methods are presented for computing this MSE, an exact method and a low-complexity approximation.

\subsection{Conventional LSQR algorithm}
LSQR is a well-known iterative algorithm for solving equalization problems of the form $\mathbf{y} = \mathbf{G}\mathbf{x} + \mathbf{w}$, where $\mathbf{x}$ is the transmitted vector, $\mathbf{y}$ is the received vector, $\mathbf{G}$ is the sparse channel matrix and  $\mathbf{w}$ is the complex AWGN noise vector with variance per dimension $\sigma^2$ \cite{Hrycak_LSQR_GMRES}. At iteration $u$, LSQR constructs the vector $\mathbf{x}_u$ in the Krylov subspace 
\begin{equation*}
    \begin{split}
        \mathcal{K}(\mathbf{G}^{\mathrm{H}}\mathbf{G},\mathbf{G}^{\mathrm{H}}\mathbf{y}, u) = \mathrm{span}\{ & \mathbf{G}^{\mathrm{H}}\mathbf{y}, (\mathbf{G}^{\mathrm{H}}\mathbf{G})\mathbf{G}^{\mathrm{H}}\mathbf{y}, \ \dots \ , \\ & (\mathbf{G}^{\mathrm{H}}\mathbf{G})^{u-1}\mathbf{G}^{\mathrm{H}}\mathbf{y} \}
    \end{split}
\end{equation*}
which minimizes the norm of the residual, $||\mathbf{y} -\mathbf{G}\mathbf{x}_k ||$. LSQR can also be regularized by including $\sigma^2$ as a damping parameter. After several iterations, LSQR provides performance similar to MMSE but with lower complexity \cite{Hrycak_LSQR_GMRES}. At each iteration, the LSQR algorithm  uses Golub-Kahan bidiagonalization and QR decomposition to obtain the estimate $\mathbf{x}_u$ \cite{Hrycak_LSQR_GMRES}. The authors of \cite{LSQR} proposed a simple recursive method for updating this estimate within each iteration. The iterative process continues until either the norm of the residual reaches a pre-determined tolerance, $\epsilon$, or the maximum number of iterations $U$ is reached. The conventional implementation of LSQR does not compute the post-equalization MSE on the symbols in $\mathbf{x}_u$ which is necessary to optimize the thresholds of the RZ detector. In order to obtain the MSE, we propose to modify the LSQR algorithm to compute this directly within the LSQR process. In the following subsections, we present two methods for computing the MSE, an exact method and a novel low-complexity approximation.

\begin{algorithm}[t]
\caption{Modified LSQR Algorithm}\label{LSQR}
\begin{algorithmic}[1]
\State \textbf{Input}: $\mathbf{G}$, $\mathbf{y}$ and $\sigma^2$
\State \textbf{Initialize}: $\mathbf{b}=\big(\begin{smallmatrix}
  \mathbf{y}\\
  \mathbf{0} \end{smallmatrix}\big)$, $\mathbf{A}=\big(\begin{smallmatrix}
  \mathbf{G}\\
  \sigma\mathbf{I} \end{smallmatrix}\big)$, $\beta_0 = \|\mathbf{b}\|$, $\mathbf{u}_0 = \mathbf{b}/\beta_0$, $\alpha_0=\|\mathbf{A}^\mathrm{H}\mathbf{u}_0\|$, $\mathbf{v}_0=\mathbf{A}^\mathrm{H}\mathbf{u}_0/\alpha_0$, $\mathbf{w}_0 = \mathbf{v}_0$, $\Bar{\phi}_0 = \beta_0$, $\Bar{\rho}_0 = \alpha_0$, $\mathbf{x}_0 = \mathbf{0}_{MN \times 1}$, $\mathbf{L}_{1} = \frac{\tau_1}{\Bar{\rho}_{0}\Bar{\phi}_{0}}\mathbf{I}_{\mathrm{MN}}$, $\mathbf{L}_0 = \mathbf{0}_{MN \times MN}$, $\tau_0=1$ and $\Bar{\phi}_u=\Bar{\rho}_u=1$ for $u<0$
\For{$u=1:U$}
  \State $\beta_u = \|\mathbf{A}\mathbf{v}_{u-1}-\alpha_{u-1}\mathbf{u}_{u-1}\|$
  \State $\mathbf{u}_u = (\mathbf{A}\mathbf{v}_{u-1}-\alpha_{u-1}\mathbf{u}_{u-1})/\beta_u$
  \State $\alpha_u = \|\mathbf{A}^\mathrm{H}\mathbf{u}_u-\beta_u\mathbf{v}_{u-1}\|$
  \State $\mathbf{v}_u = (\mathbf{A}^\mathrm{H}\mathbf{u}_u-\beta_u\mathbf{v}_{u-1})/\alpha_u$ 
  \State $\rho_u = \|[\Bar{\rho}_{u-1}\ \ \beta_u]\|$, $c_u = \frac{\Bar{\rho}_{u-1}}{\rho_u}$, $s_u = \frac{\beta_u}{\rho_u}$
  \State $\theta_u = s_u\alpha_u$, $\phi_u = c_u\Bar{\phi}_{u-1}$
  \State $\tau_u = \frac{\phi_u}{\rho_u}$, $\mu_u = \frac{\theta_u}{\rho_u}$
  \State $\Bar{\phi}_u = -s_u\Bar{\phi}_{u-1}$, $\Bar{\rho}_u = -c_u\alpha_u$
  \State $\mathbf{x}_u = \mathbf{x}_{u-1} + \tau_u\mathbf{w}_{u-1}$
  \State $\mathbf{w}_u=\mathbf{v}_u- \mu_u\mathbf{w}_{u-1}$
  \State Compute $\mathbf{L}_u$ using (\ref{eq10})
  \State if $||\mathbf{y} -\mathbf{G}\mathbf{x}_u || \leq \epsilon$, break
\EndFor
\State Compute $\psi[n]$ and $\nu[n]^2,\ $ $\forall n$ using (\ref{eq11}) and (\ref{eq12})
\State Compute $\gamma[n] = \frac{\nu[n]^2}{\psi[n]^2}, \ \forall n$
\State \textbf{Output}: $\tildebf{x} = \mathbf{x}_u$ and $\boldsymbol{\gamma}$
\end{algorithmic}
\end{algorithm}

\subsection{Exact MSE computation}
 We note that LSQR is algebraically equivalent to applying the conjugate gradient (CG) method to the normal equation $\mathbf{G}^{\mathrm{H}}\mathbf{G}\mathbf{x} = \mathbf{G}^{\mathrm{H}}\mathbf{y}$ \cite{QU_ScFDMA}. Therefore, we can adapt the method used in \cite{Yin_CG_mod} for computing the post-equalization signal-to-interference-plus-noise ratio (SINR) of the CG method to LSQR.

LSQR computes $\mathbf{x}_u$ at each iteration using a simple recursion. However, similar to the CG method in \cite{Yin_CG_mod}, $\mathbf{x}_u$ can also be computed using an LSQR equivalent equalization matrix which depends on the iteration index $u$. The LSQR equivalent equalization matrix at iteration $u$ is defined as $\mathbf{L}_u\mathbf{G}^{\mathrm{H}}$, and $\mathbf{x}_u$ can be written as
\begin{equation}
    \mathbf{x}_u = \mathbf{L}_u\mathbf{G}^{\mathrm{H}}\mathbf{y}. \label{eq6}
\end{equation}
If $\mathbf{L}_u$ is known, then the MSE on each symbol in $\mathbf{x}_u$ can be calculated. In the following, we derive a recursive method for computing $\mathbf{L}_u$ using variables which are already calculated within the LSQR process. From \cite{Yin_CG_mod}, note that the normal equation residual, $\boldsymbol{\xi}_u$, can be recursively calculated as
\begin{equation}
    \boldsymbol{\xi}_u = \boldsymbol{\xi}_{u-1} - \tau_u\mathbf{A}^\mathrm{H}\mathbf{A}\mathbf{w}_{u-1}, \label{eq7.1}
\end{equation}
where $\mathbf{A}=\big(\begin{smallmatrix}
  \mathbf{G}\\
  \sigma\mathbf{I} \end{smallmatrix}\big)$. This can also be calculated as \cite{Yin_CG_mod}
\begin{equation}
    \boldsymbol{\xi}_u = \Bar{\phi}_u\Bar{\rho}_u\mathbf{w}_u - \mu_u^2\Bar{\phi}_{u-1}\Bar{\rho}_{u-1}\mathbf{w}_{u-1}.\label{eq7}
\end{equation}
We then substitute $\boldsymbol{\xi}_u$ from (\ref{eq7}) into (\ref{eq7.1}) to obtain 
    \begin{equation}
    \begin{split}
        \Bar{\phi}_u\Bar{\rho}_u\mathbf{w}_u  = & \mu_u^2\Bar{\phi}_{u-1}\Bar{\rho}_{u-1}\mathbf{w}_{u-1} + \Bar{\phi}_{u-1}\Bar{\rho}_{u-1}\mathbf{w}_{u-1} 
        \\& - \mu_{u-1}^2\Bar{\phi}_{u-2}\Bar{\rho}_{u-2}\mathbf{w}_{u-2} - \tau_u\mathbf{A}^\mathrm{H}\mathbf{A}\mathbf{w}_{u-1}
    \end{split}
        \label{eq8}
    \end{equation}
Next, we rewrite line 12 of Algorithm 2 as $\mathbf{w}_{u-1} = (\mathbf{x}_u-\mathbf{x}_{u-1})/\tau_u$ which can then be substituted into (\ref{eq8}) to obtain the following recursion for $\mathbf{x}_u$:
    \begin{equation}
    \begin{split}
        \mathbf{x}_u  = & \mathbf{x}_{u-1} + 
        \\&\left( \frac{\tau_u\Bar{\rho}_{u-2}\Bar{\phi}_{u-2}(1+\mu_{u-1}^2)}{\tau_{u-1}\Bar{\rho}_{u-1}\Bar{\phi}_{u-1}}\mathbf{I}_{\mathrm{MN}} - \frac{\tau_u}{\Bar{\rho}_{u-1}\Bar{\phi}_{u-1}}\mathbf{A}^\mathrm{H}\mathbf{A} \right)
        \\& \times \left(\mathbf{x}_{u-1} -\mathbf{x}_{u-2}\right) 
        \\& + \frac{\mu_{u-2}^2\tau_u\Bar{\rho}_{u-3}\Bar{\phi}_{u-3}}{\tau_{u-2}\Bar{\rho}_{u-1}\Bar{\phi}_{u-1}}\left(\mathbf{x}_{u-2} -\mathbf{x}_{u-3}\right). 
    \end{split}
        \label{eq9}
    \end{equation}
Using (\ref{eq6}), we can obtain the recursion for $\mathbf{L}_u$ as
    \begin{equation}
    \begin{split}
        \mathbf{L}_u  = & \mathbf{L}_{u-1} + 
        \\&\left( \frac{\tau_u\Bar{\rho}_{u-2}\Bar{\phi}_{u-2}(1+\mu_{u-1}^2)}{\tau_{u-1}\Bar{\rho}_{u-1}\Bar{\phi}_{u-1}}\mathbf{I}_{\mathrm{MN}} - \frac{\tau_u}{\Bar{\rho}_{u-1}\Bar{\phi}_{u-1}}\mathbf{A}^\mathrm{H}\mathbf{A} \right)
        \\& \times \left(\mathbf{L}_{u-1} -\mathbf{L}_{u-2}\right) 
        \\& + \frac{\mu_{u-2}^2\tau_u\Bar{\rho}_{u-3}\Bar{\phi}_{u-3}}{\tau_{u-2}\Bar{\rho}_{u-1}\Bar{\phi}_{u-1}}\left(\mathbf{L}_{u-2} -\mathbf{L}_{u-3}\right), 
    \end{split}
        \label{eq10}
    \end{equation}
where we initialize $\mathbf{L}_{1} = \frac{\tau_1}{\Bar{\rho}_{0}\Bar{\phi}_{0}}\mathbf{I}_{\mathrm{MN}}$, $\mathbf{L}_u = \mathbf{0}_{MN \times MN}$ for $u \le 0$, $\tau_0=1$ and $\Bar{\phi}_u=\Bar{\rho}_u=1$ for $u<0$.

The matrix $\mathbf{L}_u$ can now be used to compute the MSE. Let $\mathbf{B} = \mathbf{L}_u\mathbf{Z}$, where $\mathbf{Z} = \mathbf{G}^\mathrm{H}\mathbf{G}$. The post-equalization channel gain on element $n$ of $\mathbf{x}_u$ is given by
\begin{equation}
    \psi[n] = B[n,n]. \label{eq11}
\end{equation}
The variance of the interference-plus-noise on element $n$ of $\mathbf{x}_u$ is given by 
\begin{equation}
    \nu[n]^2 = \sum_{m,m \not= n}|B[n,m]|^2 + C[n,n]\sigma^2, \label{eq12}
\end{equation}
where $\mathbf{C}=\mathbf{B}\mathbf{L}_u^\mathrm{H}.$ The MSE of element $n$ of $\mathbf{x}_u$ is therefore given by
    \begin{equation}
        \gamma[n] = \frac{\nu[n]^2}{\psi[n]^2}. \label{eq13}
    \end{equation}
 While this method provides the exact MSE of each symbol at iteration $u$ of the LSQR process, it is computationally complex due to the $MN \times MN$ matrix multiplication in (\ref{eq10}) which requires $(MN)^2$ complex multiplications. In the next subsection, we propose a approximation to this MSE which has a significantly lower computational complexity.

 \subsection{Low-complexity approximation}
 In practice, it is impossible to estimate the channel gains at each individual time sample $n$, i.e., all of the values of $h[n,l]$. Thus, we assume that the channel is varying sufficiently slowly that it has an approximately constant CIR over each OFDM symbol within an OTFS block. Under this condition, $\mathbf{G}$ is approximately a block circulant matrix with circulant blocks (BCCB) \cite{Arman_OTFS_2018}. Therefore, $\mathbf{G}$ can be diagonalized via
 \begin{equation}
     \mathbf{\Lambda_G} = (\mathbf{F}_{N}\otimes\mathbf{F}_{M})\mathbf{G}(\mathbf{F}_{N}\otimes\mathbf{F}_{M})^{\mathrm{H}}. \label{diag_G}
 \end{equation}
The matrix $\mathbf{A}^\mathrm{H}\mathbf{A}$ inherits the BCCB structure of $\mathbf{G}$. Therefore, we can obtain the diagonalization of $\mathbf{A}^\mathrm{H}\mathbf{A}$ as 
    $$ \mathbf{\Lambda_A} = (\mathbf{F}_{N}\otimes\mathbf{F}_{M})\mathbf{A}^\mathrm{H}\mathbf{A}(\mathbf{F}_{N}\otimes\mathbf{F}_{M})^{\mathrm{H}}.$$
 By using the properties of BCCB matrices \cite{Surabhi_OTFS_EQ}, we can alternatively obtain $\mathbf{\Lambda_A}$ as  
    $$\mathbf{\Lambda_A} = (\mathbf{\Lambda_G}^*\mathbf{\Lambda_G} + \sigma^2\mathbf{I}).$$
note that $\mathbf{L}_{1}$ is initialized as a diagonal matrix and hence $\mathbf{L}_{u}$ retains the BCCB structure of the $\mathbf{A}^\mathrm{H}\mathbf{A}$ for $u>2$. Additionally, since $\mathbf{L}_{1}$ is a diagonal matrix, it is invariant under diagonalization, i.e,
$$ \mathbf{\Lambda}_{\mathbf{L}_1} = (\mathbf{F}_{N}\otimes\mathbf{F}_{M}){\mathbf{L}_1}(\mathbf{F}_{N}\otimes\mathbf{F}_{M})^{\mathrm{H}} ={\mathbf{L}_1}.$$
This means that the entire recursion can be performed in the diagonalized domain. The recursion for $\mathbf{\Lambda}_{\mathbf{L}_{u}}$ can now be formulated as 
    \begin{equation}
    \begin{split}
        \mathbf{\Lambda}_{\mathbf{L}_{u}} = & \mathbf{\Lambda}_{\mathbf{L}_{u-1}} + 
        \\&\left( \frac{\tau_u\Bar{\rho}_{u-2}\Bar{\phi}_{u-2}(1+\mu_{u-1}^2)}{\tau_{u-1}\Bar{\rho}_{u-1}\Bar{\phi}_{u-1}}\mathbf{I}_{\mathrm{MN}} - \frac{\tau_u}{\Bar{\rho}_{u-1}\Bar{\phi}_{u-1}}\mathbf{\Lambda_A} \right)
        \\& \times \left(\mathbf{\Lambda}_{\mathbf{L}_{u-1}} -\mathbf{\Lambda}_{\mathbf{L}_{u-2}}\right) 
        \\& + \frac{\mu_{u-2}^2\tau_u\Bar{\rho}_{u-3}\Bar{\phi}_{u-3}}{\tau_{u-2}\Bar{\rho}_{u-1}\Bar{\phi}_{u-1}}\left(\mathbf{\Lambda}_{\mathbf{L}_{u-2}} -\mathbf{\Lambda}_{\mathbf{L}_{u-3}}\right). 
    \end{split}
        \label{eq10_apx}
    \end{equation}
where we initialize $\mathbf{\Lambda}_{\mathbf{L}_1} = \frac{\tau_1}{\Bar{\rho}_{0}\Bar{\phi}_{0}}\mathbf{I}_{\mathrm{MN}}$ and $\mathbf{\Lambda}_{\mathbf{L}_u} = \mathbf{0}_{MN \times MN}$ for $u<1$. Since this recursion only involves diagonal matrices, it can be performed with low complexity.

We can now use $\mathbf{\Lambda}_{\mathbf{L}_u}$ to calculate the approximate MSE. We calculate the diagonalizations of $\mathbf{B}$ and $\mathbf{C}$ as $\mathbf{\Lambda_B} = \tildebf{L}\mathbf{\Lambda_G}^{*}\mathbf{\Lambda_G}$ and $\mathbf{\Lambda_C} = \mathbf{\Lambda_G}^*\mathbf{\Lambda_G}\mathbf{\Lambda}_{\mathbf{L}_u}$, respectively. 
The reverse of the diagonalization process in (\ref{diag_G}) can then be used to calculate approximations of $\mathbf{B}$ and $\mathbf{C}$ as
    $$
    \tildebf{B} = (\mathbf{F}_{N}\otimes\mathbf{F}_{M})^{\mathrm{H}}\mathbf{\Lambda_B}(\mathbf{F}_{N}\otimes\mathbf{F}_{M})$$
    and
    $$\tildebf{C} = (\mathbf{F}_{N}\otimes\mathbf{F}_{M})^{\mathrm{H}}\mathbf{\Lambda_C}(\mathbf{F}_{N}\otimes\mathbf{F}_{M}).$$
Since $\tildebf{B}$ and $\tildebf{C}$ are BCCB matrices, their respective rows are simply shifted versions of each other. Therefore, under this approximation, each symbol experiences the same MSE and the subscript $n$ can be dropped from (\ref{eq11}) -- (\ref{eq13}). The post-equalization channel gain is simply given by 
\begin{equation}
\label{eq22}
    \tilde{\psi} = \Tilde{B}[1,1],
\end{equation}
and the variance of the interference-plus-noise is given by 
\begin{equation}
\label{eq23}
    \tilde{\nu}^2 = \sum_{m=2}^{MN-1}|\Tilde{B}[1,m]|^2 + \Tilde{C}[1,1]\sigma^2.
\end{equation}
Therefore, the approximate MSE on each symbol is obtained as
\begin{equation}
\label{eq24}
    \tilde{\gamma} = \frac{\tilde{\nu}^2 }{\tilde{\psi}^2}.
\end{equation}

In the context of the considered OTFS-NOMA system, we apply mLSQR in line 5 of Algorithm 1 to obtain in the $k$-th iteration the estimate of the transmitted superimposed symbol vector, $\tildebf{x}_{\mathrm{sup}}$ and the post-equalization MSE for User 1 and User 2, which is given by 
\begin{equation}
    \tilde{\gamma}_j^{(k)} = \frac{\tilde{\nu}^2 }{\rho_j\tilde{\psi}^2}, \forall j \in \{1,2\} 
\end{equation}
 We then use this calculated MSE to optimize the thresholds of the RZ detector in a greedy manner, as described in detail in the following section. 

It is important to note that the proposed modifications to the LSQR algorithm do not change the computational procedure of LSQR; instead, the modifications use terms that are already calculated in LSQR to obtain the post-equalization MSE. As such, the proposed modifications do not affect the numerical stability of LSQR.

\section{RZ Detector Threshold Optimization}
In this section, we describe how the MSE calculated by the modified LSQR algorithm can be used to optimize the RZ thresholds for each user. Our proposed method works by tracking the evolution of the MSE on the symbols of User 1 and User 2 as Algorithm 1 progresses. The key idea is to choose optimal values for the RZ thresholds $T_1$ and $T_2$ in each iteration $k$ which minimise the ``pre-decision" MSE, i.e., the mean-square value of the error seen by the RZ detector at iteration $k+1$. 

In lines 15 and 16 of Algorithm 1, at iteration $k$, the RZ detector makes a decision on whether the received symbols of each user are unreliable or reliable and then quantizes the reliable symbols to the nearest QAM symbol in that user's constellation. Therefore, there are 3 possible outcomes of the unreliable zone detection. Symbols are either correct, incorrect or undetected, each such event having its own associated probability which depends on the thresholds, $T_1$ and $T_2$, and the user's MSE values, $\tilde{\gamma}_1^{k}$ and $\tilde{\gamma}_2^{k}$.

Next, we derive expressions for the probability of each outcome above in the context of each user's symbols. To derive the probability of each outcome for a generalized $A_i$-ary QAM system, we first derive them for a $\sqrt{A_i}$-ary PAM system by adapting the closed-form expression for the probability of error of a 2-user NOMA system derived in \cite{He_NOMA_SER_2019}. For the User 1 symbols, the decision is being made on the superimposed symbols which contain contributions from the symbols of User 1 \textit{and} User 2. First, the following functions are defined (c.f.\cite{He_NOMA_SER_2019}):
\begin{equation*}
        \mathrm{q_a}(j,l,T_1,\tilde{\gamma}_1) = \mathrm{Q}\left(\frac{d((2j-1)-(2l-1)\sqrt{\frac{\rho_2}{\rho_1}})-\frac{T_1}{2}}{\sqrt{\tilde{\gamma}_1/2}}  \right),
\end{equation*}
    \begin{equation*}
        \mathrm{q_b}(j,l,T_1,\tilde{\gamma}_1) = \mathrm{Q}\left(\frac{d((2j-1)+(2l-1)\sqrt{\frac{\rho_2}{\rho_1}})-\frac{T_1}{2}}{\sqrt{\tilde{\gamma}_1/2}}  \right),
    \end{equation*}
    \begin{equation*}
        \mathrm{q_c}(l,T_1,\tilde{\gamma}_1) = \mathrm{Q}\left(\frac{d(1-(2l-1)\sqrt{\frac{\rho_2}{\rho_1}})+\frac{T_1}{2}}{\sqrt{\tilde{\gamma}_1/2}}  \right),
    \end{equation*}
    \begin{equation*}
        \mathrm{q_d}(l,T_1,\tilde{\gamma}_1) = \mathrm{Q}\left(\frac{d(1+(2l-1)\sqrt{\frac{\rho_2}{\rho_1}})+\frac{T_1}{2}}{\sqrt{\tilde{\gamma}_1/2}}  \right),
    \end{equation*}
where  $\mathrm{Q}(x) = \frac{1}{2}\mathrm{erfc}(\frac{x}{\sqrt{2}})$ denotes the Gaussian Q-function. The probability of correct symbol detection per dimension for User 1, denoted by $P_{\mathrm{C,PAM},1}$, is then given by (\ref{eq29}), shown at the top of the next page, where the threshold used is $T_1$. For the probability of incorrect detection, we adopt a \textit{nearest-neighbor approximation}, i.e., it is assumed that if an incorrect symbol is detected, it is always a nearest neighbor in that user's QAM constellation (this assumption becomes very accurate at high SNR). The probability of incorrect detection per dimension for User 1, denoted by $P_{\mathrm{E,PAM},1}$, is given by (\ref{eq30}), where the threshold used is $T_1$.
\begin{figure*}[t]
    \normalsize
    \setcounter{equation}{30}
    \begin{equation}
    \label{eq29}
        P_{\mathrm{C,PAM},1}=  1 - \frac{2(\sqrt{A_1}-1)}{A_1}\sum_{l=1}^{\sqrt{A_1}/2}\left[\mathrm{q_a}(1,l,T_1,\tilde{\gamma}_1) +\mathrm{q_b}(1,l,T_1,\tilde{\gamma}_1)\right],
    \end{equation}
    
    \begin{equation}
    \begin{split}
    \label{eq30}
        P_{\mathrm{E,PAM},1}&=  \frac{2}{A}\left(\sum_{l=1}^{\sqrt{A_1}/2}\left[(\sqrt{A_1}-1)\left(\mathrm{q_c}(1,l,T_1,\tilde{\gamma}_1) +\mathrm{q_d}(1,l,T_1,\tilde{\gamma}_1)\right) - (\sqrt{A_1}-2)\left(\mathrm{q_a}(2,l,T_1,\tilde{\gamma}_1) +\mathrm{q_b}(2,l,T_1,\tilde{\gamma}_1)\right)\right] \right)
        \end{split}
    \end{equation}
    
    \begin{equation}
    \label{eq31}
        P_{\mathrm{C,PAM},2}=  1 - \frac{2(\sqrt{A_2}-1)}{\sqrt{A_2}}\mathrm{Q}\left(\frac{d-\frac{T_2}{2}}{\sqrt{\tilde{\gamma}_2/2}}  \right),
    \end{equation}
    
    \begin{equation}
    \begin{split}
    \label{eq32}
        P_{\mathrm{E,PAM},2}&=  \frac{2}{\sqrt{A_2}}\left((\sqrt{A_2}-1)\mathrm{Q}\left(\frac{d+\frac{T_2}{2}}{\sqrt{\tilde{\gamma}_2/2}}  \right) - (\sqrt{A_2}-2)\mathrm{Q}\left(\frac{3d-\frac{T_2}{2}}{\sqrt{\tilde{\gamma}_2/2}}  \right) \right)
        \end{split}
    \end{equation}
\hrulefill
\vspace*{4pt}
\end{figure*}
User 2 symbols are only fed into the RZ detector once the corresponding User 1 symbols have been detected on a previous iteration. Thereafter, the decisions are no longer being made upon a superposition of both user symbols. Consequently, the probability of correct detection per dimension of User 2 is given by (\ref{eq31}), where the threshold used is $T_2$. The probability of incorrect detection per dimension of User 2 is given by (\ref{eq32}). The probability of correct detection, incorrect detection, and non-detection at User $i \in \{1,2 \}$ can then be obtained, respectively, as
    \begin{equation}
        P_{\mathrm{c},i} = (P_{\mathrm{C,PAM},i})^2, \label{eq33}
    \end{equation}
    \begin{equation}
        P_{\mathrm{e},i} = 2P_{\mathrm{E,PAM},i}, \label{eq34}
    \end{equation}
%
%
    \begin{equation}
        P_{\mathrm{u},i} = 1-P_{\mathrm{c},i} - P_{\mathrm{e},i}, \label{eq36}
    \end{equation}
The next subsection describes how these probability expressions can be used to optimize the thresholds at each user's receiver. 

\subsection{Design of threshold $T_i$ at User $i$ receiver}
In this subsection we describe the process for choosing the threshold $T_i$ at the receiver of User $i$ in each iteration of Algorithm 1. We begin at iteration $k=1$ where it can be seen from (\ref{eq22}), (\ref{eq23}) and (\ref{eq24}) the MSE of the User $i$ symbols after mLSQR equalization is given by 
\begin{equation}
    \tilde{\gamma}_i^{(1)}= \frac{1}{\rho_i\tilde{\psi}^2}\left(\sum_{m=2}^{MN-1}\!\!\!\!|\Tilde{B}[1,m]|^2(\rho_1 + \rho_2) + \Tilde{C}[1,1]\sigma^2\right). \label{eq38}
\end{equation}
%
%
This can be rewritten as
\begin{equation}
        \tilde{\gamma}_i^{(1)} = \Omega_i^{(1)} + \Psi_{i,u}^{(1)} + W_i. 
        \label{eq39}
\end{equation}
where $\Omega_i^{(1)} = \frac{\rho_1}{\rho_i\tilde{\psi}^2}\sum_{m=2}^{MN-1}|\Tilde{B}[1,m]|^2$ is the MSE due to the undetected User 1 symbols, $\Psi_{i,u}^{(1)} = \frac{\rho_2}{\rho_i\tilde{\psi}^2}\sum_{m=2}^{MN-1}|\Tilde{B}[1,m]|^2$ is the MSE due to the undetected User 2 symbols and $W_i$ is the AWGN component of the MSE. After the detection and interference cancellation process in lines 15--17 of Algorithm 1, the MSE due to undetected User 1 symbols will be reduced by a factor depending on the probability of non-detection of User 1 symbols in iteration 1. Therefore, we can express the remaining MSE of the undetected User 1 symbols at iteration $k=2$ as $\Omega_i^{(2)} = \Omega_i^{(1)}P_{\mathrm{u},1}^{(1)}$. Generalizing this argument, at iteration $k$, we express the remaining MSE of the undetected User 1 symbols as $\Omega_i^{(k)} = \Omega_i^{(k-1)}P_{\mathrm{u},1}^{(k-1)}$ and we define the remaining MSE of the undetected User 2 symbols as $\Psi_{i,u}^{(k)} = \Psi_{i,u}^{(k-1)}P_{\mathrm{u},2}^{(k-2)}$. Since $W_i$ is unaffected by the interference cancellation process, it sets a limit on the minimum achievable probability of error. Therefore, the User $i$ receiver should choose a threshold at iteration $k$ which achieves this minimum minimum achievable probability of error. We note that, via (\ref{eq34}), (\ref{eq30}) and (\ref{eq32}), $ P_{\mathrm{e},i}$ can be expressed as a function of 2 variables, i.e, $\tilde{\gamma}_i$ and $T_i$. Hence,  the User $i$ receiver chooses the threshold $T_i$ at iteration $k$ such that
\begin{equation}
    P_{\mathrm{e},i}(\tilde{\gamma}_i^{(k)}, T_i) = P_{\mathrm{e},i}(W_i, 0). \label{eq40}
\end{equation}
Since the remaining User $j$ symbols $(j \ne i)$ impart MUI on the remaining undetected User $i$ symbols, User $i$ must select the threshold $T_j$ which minimizes $\tilde{\gamma}_i^{(k)}$. We now describe the exact optimization process at each user's receiver.

\subsection{Optimizing $T_2$ at receiver of User 1}

\begin{figure*}[t]
        \normalsize
        \setcounter{equation}{49}

    \begin{equation}
        \label{eq46}
        \begin{split}
        \frac{\mathrm{\partial}}{\mathrm{\partial}T_1}P_{\mathrm{c},1}^{(k)} =& \left(2-\frac{4(\sqrt{A_1}-1)}{A_1}\sum_{l=1}^{\sqrt{A_1}/2}\left[\mathrm{q_a}(1,l) +\mathrm{q_b}(1,l)\right]\right) \times
        \\& \left(\frac{(\sqrt{A_1}-1)}{A_1\sqrt{\pi\tilde{\gamma}_1^{(k)}}}\sum_{l=1}^{\sqrt{A_1}/2}\left[
        \exp\left(-\frac{(d((4l-2)\sqrt{\frac{\rho_2}{\rho_1}}-2)+{T_1})^2}{4\tilde{\gamma}_1^{(k)}}\right) 
        + \exp\left(-\frac{(d((2-4l)\sqrt{\frac{\rho_2}{\rho_1}}-2)+{T_1})^2}{4\tilde{\gamma}_1^{(k)}}\right) \right]\right)
        \end{split}
    \end{equation}

        \begin{align}
        \label{eq47}
        \frac{\mathrm{\partial}}{\mathrm{\partial}T_1}P_{\mathrm{e},1}^{(k)} =& \frac{-2}{A_1\sqrt{\pi\tilde{\gamma}_1^{(k)}}} \times \nonumber\\
        & \left(\sum_{l=1}^{\sqrt{A_1}/2}\left[(\sqrt{A_1}-1)\left(\exp\left(-\frac{(d((2-4l)\sqrt{\frac{\rho_2}{\rho_1}}+2)+{T_1})^2}{4\tilde{\gamma}_1^{(k)}}\right)
        + \exp\left(-\frac{(d((4l-2)\sqrt{\frac{\rho_2}{\rho_1}}+2)+{T_1})^2}{4\tilde{\gamma}_1^{(k)}}\right)\right) \nonumber \right. \right.\\
        & \left.\left. +(\sqrt{A_1}-2)\left(\exp\left(-\frac{(d((4l-2)\sqrt{\frac{\rho_2}{\rho_1}}-6)+{T_1})^2}{4\tilde{\gamma}_1^{(k)}}\right)
        +\exp\left(-\frac{(d((2-4l)\sqrt{\frac{\rho_2}{\rho_1}}-6)+{T_1})^2}{4\tilde{\gamma}_1^{(k)}}\right)\right)\right] \right)
        \end{align}

    \begin{align}
    \label{eq48}
        \frac{\mathrm{\partial}}{\mathrm{\partial}T_2}P_{\mathrm{c},2}^{(k)} =& \left(2-\frac{4(\sqrt{A_2}-1)}{A_2}\mathrm{Q}\left(\frac{d-\frac{T_2}{2}}{\sqrt{\tilde{\gamma}_2^{(k)}/2}}\right)\right)\left(\frac{(\sqrt{A_2}-1)}{A_2\sqrt{\pi\tilde{\gamma}_2^{(k)}}} \exp\left(-\frac{(2d+{T_2})^2}{4\tilde{\gamma}_2^{(k)}}\right)\right)
    \end{align}

    \begin{align}
    \label{eq49}
        \frac{\mathrm{\partial}}{\mathrm{\partial}T_2}P_{\mathrm{e},2}^{(k)} =& \frac{-1}{A_2\sqrt{\pi\tilde{\gamma}_1^{(k)}}}\left((\sqrt{A_2}-1)\exp\left(-\frac{(2d+{T_2})^2}{4\tilde{\gamma}_2^{(k)}}\right)
        + (\sqrt{A_2}-2)\exp\left(-\frac{({T_2}-6d)^2}{4\tilde{\gamma}_2^{(k)}}\right) \right)
    \end{align}
    \hrulefill
    \vspace*{4pt}
\end{figure*}

\setcounter{equation}{40}

The MSE of User 1 will be reduced by the correctly detected symbols from the previous iteration and increased by the incorrectly detected symbols. Hence, the MSE of User 1 at iteration 2 will be comprised of the remaining interference from the undetected User 1 symbols, the interference from the undetected User 2 symbols, the AWGN and the MSE due to interference cancellation error multiplied by the probability of error of User 1. Therefore, the MSE for User 1 at iteration 2 is given by
\begin{align}
    \tilde{\gamma}_1^{(2)} &= \Omega_1^{(1)}P_{\mathrm{u},1}^{(1)} + E_1\Omega_1^{(1)}P_{\mathrm{e},1}^{(1)} + \Psi_{1,u}^{(1)} + W,
        \label{eq41}
\end{align}
where $E_i = 4d_i^2$ is the average energy of an interfering symbol due to the event of interference cancellation error of User $i$ under the nearest-neighbour approximation. This is multiplied by the probability of error of User $i$ and by the remaining MSE due to undetected User $i$ symbols to account for the reduced number of symbols can be incorrectly detected as Algorithm 1 progresses.  Using (\ref{eq36}), we can express (\ref{eq41}) as
\begin{equation}
    \tilde{\gamma}_1^{(2)} = \tilde{\gamma}_1^{(1)} - \Omega_1^{(1)}P_{\mathrm{c},1}^{(1)} + (E_1-1)\Omega_1^{(1)}P_{\mathrm{e},1}^{(1)}, 
        \label{eq42}
\end{equation}
 Generalizing this argument, we can formulate an expression for the MSE of User 1 on iteration $k+1$, which is given by
    \begin{equation}
    \begin{split}
        \tilde{\gamma}_1^{(k+1)} &= \tilde{\gamma}_1^{(k)} - \Omega_1^{(k)}P_{\mathrm{c},1}^{(k)} + (E_1-1)\Omega_1^{(k)}P_{\mathrm{e},1}^{(k)}
        \\& -\left(\Psi_{1,u}^{(k)}\left(P_{\mathrm{c},1}^{(k-1)} + P_{\mathrm{e},1}^{(k-1)}\right) + \Psi_{1,d}^{(k)}\right)P_{\mathrm{c},2}^{(k)}
        \\& + (E_2-1)\left(\Psi_{1,u}^{(k)}\left(P_{\mathrm{c},1}^{(k-1)} + P_{\mathrm{e},1}^{(k-1)}\right) + \Psi_{1,d}^{(k)}\right)P_{\mathrm{e},2}^{(k)}. 
        \label{eq43}
    \end{split}
    \end{equation} 
where $\Psi_{1,d}^{(k)} = \Psi_{1,d}^{(k-1)} + \Psi_{1,u}^{(k-1)}\left(P_{\mathrm{c},1}^{(k-2)} + P_{\mathrm{e},1}^{(k-2)}\right)P_{\mathrm{u},2}^{(k-1)} - \Psi_{1,d}^{(k-1)}\left(P_{\mathrm{c},2}^{(k-1)} + P_{\mathrm{e},2}^{(k-1)}\right)$ is the remaining interference power from the User 2 symbols for which the corresponding User 1 symbols have been detected. 

The probability of User 1 symbols being undetected is initialized as $P_{\mathrm{u},1}^{(0)}=1$. We also initialize $\Psi_{1,d}^{(-1)}=0$ as no User 1 symbols have been detected before the algorithm begins. The MSE for User 1 on iteration $k+1$ is a function of the probability terms in (\ref{eq33}) and (\ref{eq34}), which are themselves functions of $T_2$. All other terms are constants which can be updated recursively. The User 1 receiver can now choose the optimum value $T_2$ at iteration $k$ to minimize the MSE of User 1 at iteration $k+1$ .

At each iteration $k$, the receiver of User 1 solves the optimization problem
    \begin{subequations}
        \begin{alignat}{2}
            &\min_{T_2}    &\qquad& \tilde{\gamma}_{1}^{(k+1)} \label{eq44}\\
            &\text{s.t.} &      & T_2 \geq 0 .\label{eq:constraint1}
        \end{alignat}
    \end{subequations}
To solve this optimization problem, the derivative of $\tilde{\gamma}_{1}^{(k+1)}$ with respect to $T_2$ is set equal to zero. Since only $P_{\mathrm{e},2}^{(k)}$ and $P_{\mathrm{c},2}^{(k)}$ in (\ref{eq43}) are functions of $T_2$, the derivative of $\tilde{\gamma}_{1}^{(k+1)}$ with respect to $T_2$ is given by 
\begin{equation}
    \begin{split}
        \frac{\mathrm{\partial}}{\mathrm{\partial}T_2}\tilde{\gamma}_{1}^{(k+1)} =  &\left(\Psi_{2,u}^{(k)}P_{\mathrm{d},1}^{(k-1)} + \Psi_{1,d}^{(k)}\right) \times
        \\& \left((E_2-1)\frac{\mathrm{\partial}}{\mathrm{\partial}T_2}P_{\mathrm{e},2}^{(k)}
        -\frac{\mathrm{\partial}}{\mathrm{\partial}T_2}P_{\mathrm{c},2}^{(k)}\right). 
        \end{split}\label{eq45}
    \end{equation}
Using $\frac{\mathrm{\partial}}{\mathrm{\partial}x}\mathrm{Q}(x) = -\frac{1}{\sqrt{2\pi}}e^{-x^2}$, the derivative of $P_{\mathrm{c},2}^{(k)}$ and $P_{\mathrm{e},2}^{(k)}$ can be expressed as (\ref{eq48}) and (\ref{eq49}), respectively. The User 1 receiver then solves  $\frac{\mathrm{\partial}}{\mathrm{\partial}T_2}\tilde{\gamma}_{1}^{(k+1)} = 0$ using the Brent-Dekker method \cite{brent71:CJ-14-422} to obtain the solution to (\ref{eq44}), which is the optimized $T_2$ at iteration $k$ of Algorithm 1.

\subsection{Optimizing $T_1$ at  User 2 receiver}

As with User 1, the MSE of User 2 will be reduced by the correctly detected symbols from the previous iteration and increased by the incorrectly detected symbols. However, in contrast to User 1, the RZ detector of User 2 only makes decisions on the User 2 symbols whose corresponding User 1 symbols have already been detected. Therefore, the MSE of a User 2 symbol is also affected by the incorrect detection of the overlapping User 1 symbol. Given this, at iteration 2 of Algorithm 1, the MSE of the User 2 symbols which are being fed into the RZ detector can be written as 
\begin{equation}
\begin{split}
        \tilde{\gamma}_2^{(2)} = & \tilde{\gamma}_2^{(1)} - \Omega_2^{(1)}P_{\mathrm{c},1}^{(1)} + (E_1-1)\Omega_2^{(1)}P_{\mathrm{e},1}^{(1)} 
        \\& + \frac{\rho_2}{\rho_1}E_1 P_{\mathrm{u},1}^{(0)}P_{\mathrm{e},1}^{(1)} + W, \label{eq50}
\end{split}
\end{equation}
where the fourth term of (\ref{eq50}) accounts for the MSE due to directly overlapping User 1 symbols that are incorrectly detected. Generalizing this argument, we can formulate a general expression for the MSE of User 2 at iteration $k+1$, which is given by
    \begin{equation}
    \begin{split}
        \tilde{\gamma}_2^{(k+1)} = &\tilde{\gamma}_2^{(k)} - \Omega_2^{(k)}P_{\mathrm{c},1}^{(k)}
        \\& +((E_1-1)\Omega_2^{(k)}+\frac{\rho_2}{\rho_1}E_1 P_{\mathrm{u},1}^{(1)})P_{\mathrm{e},1}^{(k)}
        \\& -\left(\Psi_{2,u}^{(k)}P_{\mathrm{d},1}^{(k-1)} + \Psi_{2,d}^{(k)}\right)P_{\mathrm{c},2}^{(k)}
        \\& + (E_2-1)\left(\Psi_{2,u}^{(k)}P_{\mathrm{d},1}^{(k-1)} + \Psi_{2,d}^{(k)}\right)P_{\mathrm{e},2}^{(k)}. 
        \label{eq51}
    \end{split}
    \end{equation} 
Similar to the case of the User 1 receiver above, we can now formulate the optimization problem to be solved at iteration $k$, i.e.,
    \begin{subequations}
        \begin{alignat}{2}
            &\min_{T_1}    &\qquad& \tilde{\gamma}_{2}^{(k+1)} \label{eq52}\\
            &\text{s.t.} &      & T_1 \geq 0 ,\label{eq:constraint2}
        \end{alignat}
    \end{subequations}
This optimization problem is solved in a similar manner to (\ref{eq44}). Since only $P_{\mathrm{e},1}^{(k)}$ and $P_{\mathrm{c},1}^{(k)}$ in (\ref{eq24}) are functions of $T_1$, the derivative of $\tilde{\gamma}_{2}^{(k+1)}$ with respect to $T_1$ is given by 
\begin{equation}
    \begin{split}
        \frac{\mathrm{\partial}}{\mathrm{\partial}T_1}\tilde{\gamma}_{1}^{(k+1)} = &  \left((E_1-1)\Omega_2^{(k)}+\frac{\rho_2}{\rho_1}EP_{\mathrm{u},1}^{(1)}\right)\frac{\mathrm{\partial}}{\mathrm{\partial}T_1}P_{\mathrm{e},1}^{(k)}
        \\& -\Omega_2^{(k)}\frac{\mathrm{\partial}}{\mathrm{\partial}T_1}P_{\mathrm{c},1}^{(k)} 
        \end{split}\label{eq53}
    \end{equation}
The derivatives of $P_{\mathrm{c},2}^{(k)}$ and $P_{\mathrm{e},2}^{(k)}$ can be expressed as (\ref{eq46}) and (\ref{eq47}) respectively. The receiver of User 2 then solves  $\frac{\mathrm{\partial}}{\mathrm{\partial}T_2}\tilde{\gamma}_{1}^{(k+1)} = 0$ to obtain the solution to (\ref{eq52}), which is the optimized $T_1$ at iteration $k$ of Algorithm 1.

\subsection*{Computational complexity}
In this subsection, the computational complexity of the proposed method is compared to that of the MMSE-SIC benchmark, in terms of the number of complex multiplications. Direct implementation of MMSE equalization involves the inversion of an $MN \times MN$ matrix and hence has a computational complexity of $\mathcal{O}(M^3N^3)$. Each iteration of the conventional LSQR algorithm has a computational complexity of $\mathcal{O}(MN\log_2(MN))$ \cite{Hrycak_LSQR_GMRES}. The low-complexity MSE calculation in the proposed mLSQR algorithm (described in Subsection IV-C) can be performed with a single $M$-point FFT operation and a single $N$-point IFFT operation and therefore has a computational complexity of $\mathcal{O}(M\log_2(M)) + \mathcal{O}(N\log_2(N))$, which is negligible compared to the complexity of the LSQR computation. Hence, the proposed mLSQR algorithm has a computational complexity of $\mathcal{O}(MN\log_2(MN))$. In the worst-case scenario, Algorithm 1 performs mLSQR $K$ times, each with $U$ mLSQR iterations; therefore, the computational complexity of Algorithm 1 is $\mathcal{O}(UKMN\log_2(MN))$. In practice, the typical values of $K$ and $U$ are in the order of tens and the typical values of $M$ and $N$ can be as high as $M=512$ and $N=128$ \cite{Tiwar_OTFS_LMMSE, Rav_INT_canc_MP}. Thus, $UK \ll M^2N^2$ and our method can achieve orders of magnitude computational complexity improvement over MMSE-SIC for OTFS-NOMA. It should also be noted that optimizing the thresholds allows for Algorithm 1 to converge faster at high SNR than a naive threshold design, as the thresholds are not unnecessarily large and the algorithm can detect more symbols at earlier iterations.

\section{Numerical Results and Discussion}

\begin{figure}[t]
    \centering
    \includegraphics[width = \columnwidth]{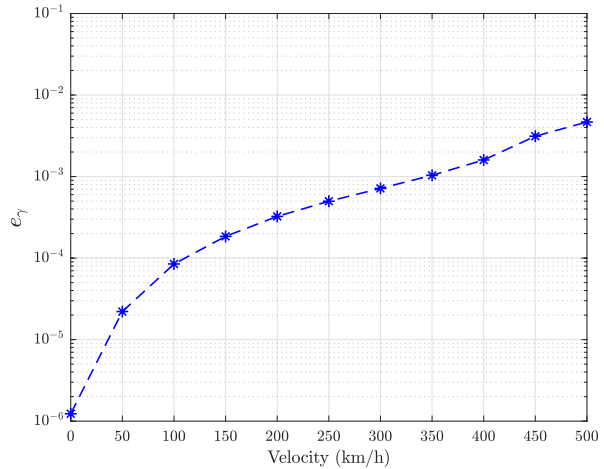}
    \caption{Normalized approximation error of the proposed low-complexity MSE computation.}
    \label{fig:mlsqr_apx_err}
\end{figure}

This section presents numerical results to showcase the effectiveness of the proposed OTFS-NOMA equalization and detection algorithm. As a benchmark, an OTFS-NOMA system using MMSE equalization and SIC for detection is considered, which is referred to as MMSE-SIC. Additionally, the performance of the proposed algorithm using the optimized thresholds outlined in Section~V is compared to the proposed algorithm with naive (conventional) threshold design. For the naive threshold case, we consider a starting threshold of $T_i^{(1)} = 2d_i$ for each user which is then reduced geometrically within each iteration as $T_i^{(k)} = T_i^{(1)}(1-(k/K))$ (this was the threshold adaptation strategy adopted in \cite{Qu_OTFS_detection, Hampeis_LSQR, Taubock_LSQR_2011}). Monte Carlo simulation is used to average the results over $10^5$ random channel instances. 

\begin{table}[h]
    \caption{Simulation Parameters}
    \centering
    \begin{tabular}{|c|c|}
        \hline
         Delay bins ($M$) & 64\\
         \hline
         Doppler bins ($N$) & 16 \\
         \hline
         Carrier frequency ($f_c$) & 5.9 GHz \\
         \hline
         Subcarrier spacing  & 15 kHz \\
         \hline
         Modulation scheme & 4-QAM, 16-QAM  \\
         \hline
         Channel model & TDL-C \cite{3gpp_TS38901} \\
         \hline
         Delay spread & 300 ns  \\
         \hline
         User velocity & $90 - 450$ km/h  \\
         \hline
         Algorithm 1 iterations ($K$)  & 10  \\
         \hline
         mLSQR iterations ($U$)  & 15  \\
         \hline
         mLSQR tolerance ($\epsilon$)  & $10^{-2}$  \\
         \hline
    \end{tabular}
    \label{tab:table 1}
\end{table}

\begin{figure}[t]
    \centering
    \includegraphics[width = \columnwidth]{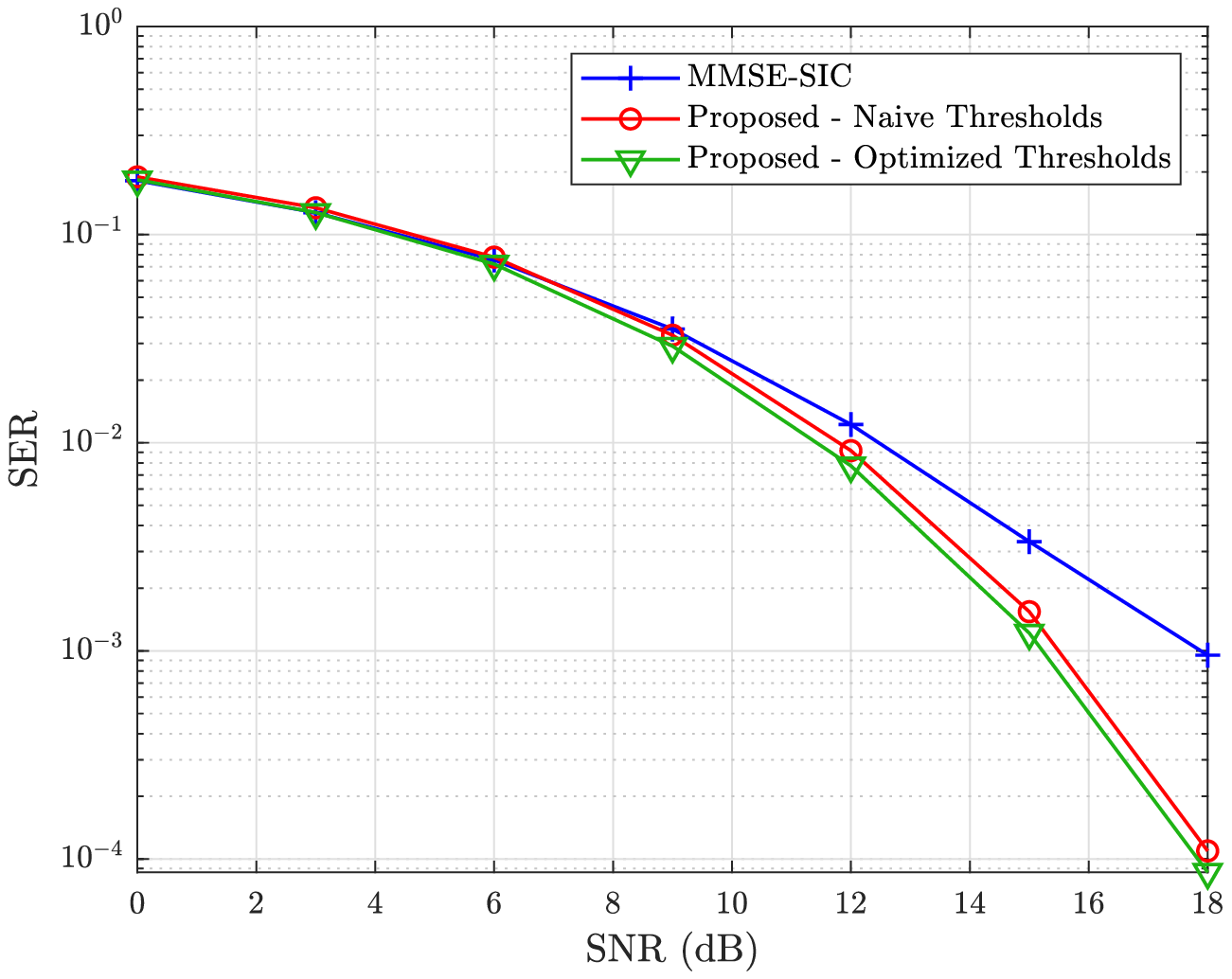}
    \caption{Comparison of the SER performance of User 1 using Algorithm 1 with optimized thresholds, Algorithm 1 with naive thresholds, and MMSE equalization with SIC, with different SNR levels, for the case where each user is allocated a 4-QAM constellation.}
    \label{fig:U1_SNR}
\end{figure}

\begin{figure}[t]
    \centering
    \includegraphics[width = \columnwidth]{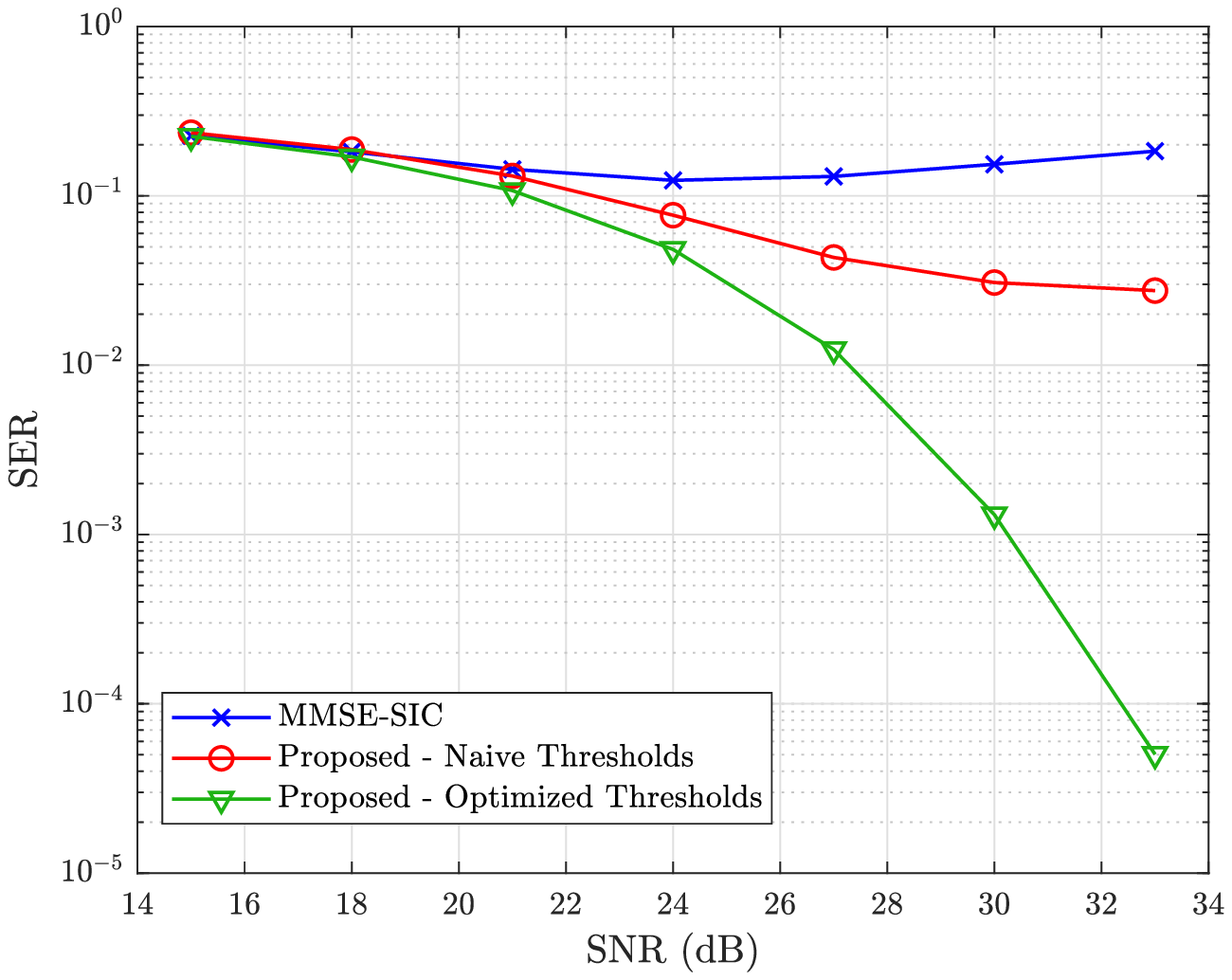}
    \caption{Comparison of the SER performance of User 2 using Algorithm 1 with optimized thresholds, Algorithm 1 with naive thresholds, and MMSE equalization with SIC, with different SNR levels, for the case where each user is allocated a 4-QAM constellation.}
    \label{fig:U2_SNR}
\end{figure}

\begin{figure}[t]
    \centering
    \includegraphics[width = \columnwidth]{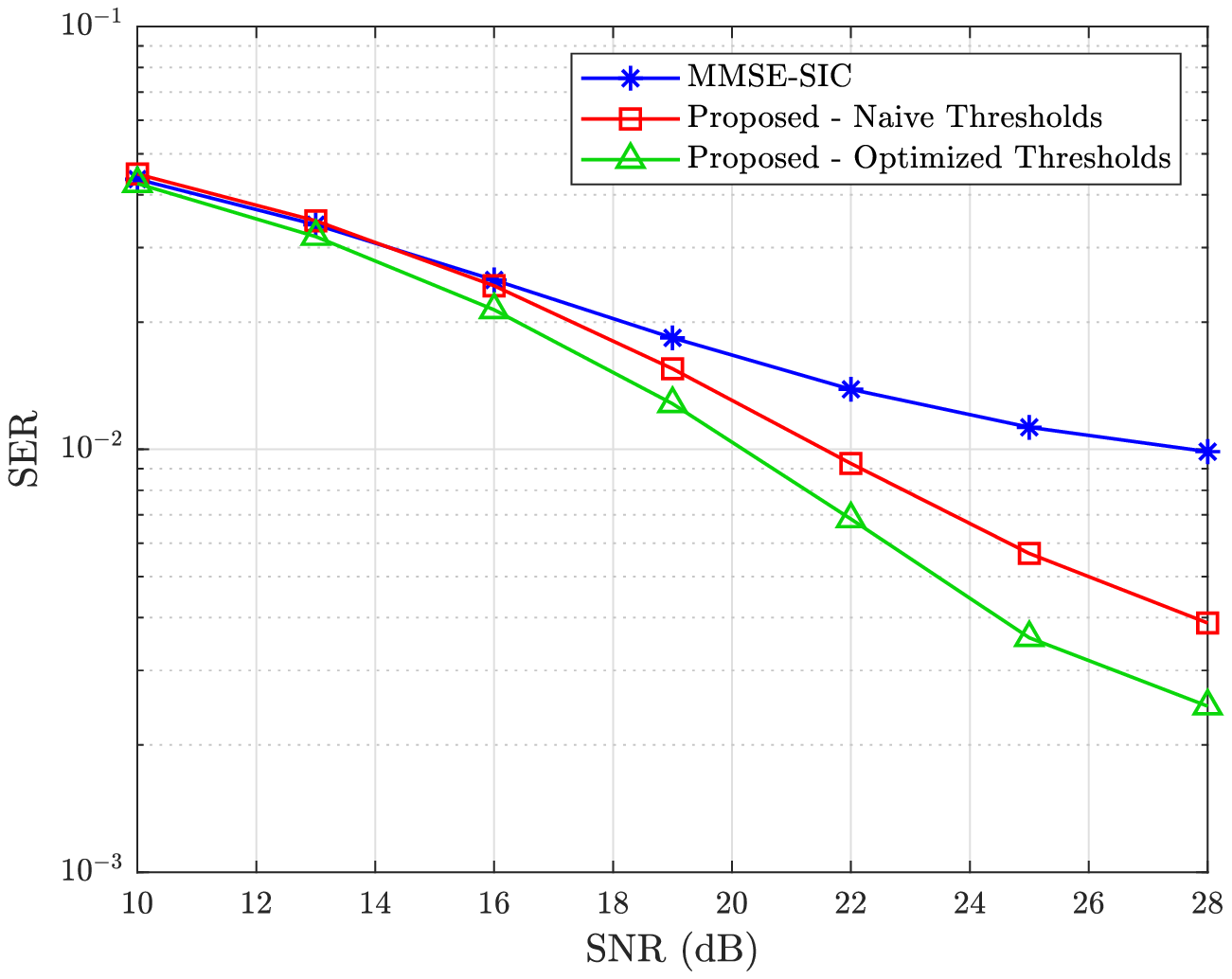}
    \caption{Comparison of the SER performance of User 1 using Algorithm 1 with optimized thresholds, Algorithm 1 with naive thresholds, and MMSE equalization with SIC, with different SNR levels, for the case where each user is allocated a 16-QAM constellation.}
    \label{fig:U1_SNR_16}
\end{figure}

\begin{figure}[t]
    \centering
    \includegraphics[width = \columnwidth]{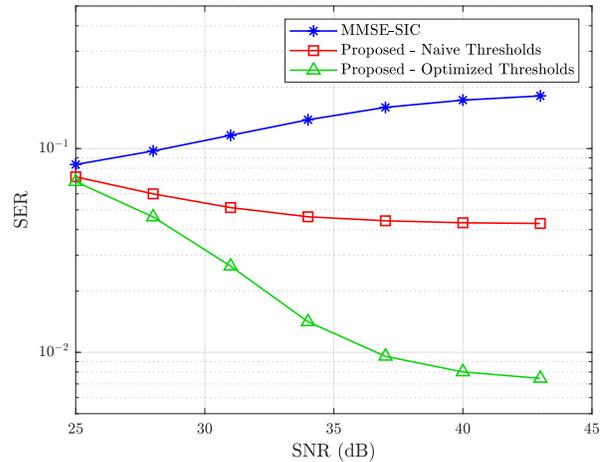}
    \caption{Comparison of the SER performance of User 2 using Algorithm 1 with optimized thresholds, Algorithm 1 with naive thresholds, and MMSE equalization with SIC, with different SNR levels, for the case where each user is allocated a 16-QAM constellation.}
    \label{fig:U2_SNR_16}
\end{figure}

A carrier frequency of $f_c = 5.9$ GHz, a transmission bandwidth of 4.95~MHz and a delay-Doppler grid size of $M=64$ and $N=16$ are considered. Additionally, we consider a fixed SNR difference of 15~dB between the users, i.e., User 2 has an average SNR that is 15~dB higher than that of User 1. The Tapped Delay Line C (TDL-C) model with a delay spread of 300~ns \cite{3gpp_TS38901} is used for the channel model. We consider a range of maximum Doppler shifts from 500~Hz to 2500~Hz, which corresponds to velocities of approximately 90~km/h to 450~km/h at a carrier frequency of 5.9 GHz. The Doppler shifts are generated using Jakes' model \cite{jakes_model}.  For the mLSQR algorithm, a maximum number of iterations of $U=15$ and a tolerance of $\epsilon = 10^{-2}$ are used, which are commonly used values for LSQR implementation in the related literature \cite{Qu_OTFS_detection, QU_ScFDMA}. Additionally, the (low-complexity) approximate MSE computation method outlined in subsection~IV-C is used in the mLSQR algorithm for all simulations. For Algorithm 1, we consider a maximum number of iterations of $K=10$ to limit the computational complexity. For power allocation, we use the average-SNR-based fractional transmit power allocation (FTPA) scheme outlined in \cite{Chatt_OTFS_NOMA}. The scheme works by considering the average SNR of each user as a fraction of the sum of the SNR of both users. The transmit power of User $i$ is given by:
$$ \rho_i = \frac{\mathrm{SNR}_i}{\mathrm{SNR}_1 + \mathrm{SNR}_2}.$$

To compare the low-complexity MSE computation outlined in Section~IV-C to the exact method outlined in Section~IV-B, we demonstrate the approximation error of the low-complexity method. We define the normalized approximation error as 
\begin{equation}
    e_\gamma = \frac{1}{MN}\mathbb{E}\left\{\sum_{n=0}^{MN-1}|{\gamma}_n - \tilde{\gamma}|^2\right\}
\end{equation}
Figs. \ref{fig:mlsqr_apx_err} shows $e_\gamma$ at different velocities with an SNR of 15~dB.  For this simulation, a small scale example is considered, where $M=N=4$, due to the computational complexity of the exact MSE computation method. As can be seen in Fig. \ref{fig:mlsqr_apx_err}, the approximation error is very small at low velocities, which demonstrates the validity of the low-complexity method when the channel matrix structure is close to BCCB. As expected, the error becomes larger as the velocity increases, as the assumption of a BCCB channel matrix becomes less valid. However, the approximation error is still relatively small and the low-complexity approximate MSE calculation is still useful for choosing the user thresholds. 
 
Fig. \ref{fig:U1_SNR} shows the symbol error rate (SER) of User 1 using the proposed equalization and detection method compared to the benchmark schemes for different signal-to-noise ratio (SNR) conditions. For these simulations each user's symbols are taken from a 4-QAM constellation, i.e., $A_1 = A_2 = 4$, and the user velocity is fixed at 200 km/h, which equates to a maximum Doppler shift of approximately 1000~Hz. It can be seen from Fig. \ref{fig:U1_SNR} that for User 1, the proposed method outperforms the MMSE-SIC method, providing an SNR gain more than 2~dB at an SER of $10^{-3}$. Additionally, optimizing the RZ detector thresholds provides further performance gains over the naive threshold design benchmark. Since User 1 has a larger power allocation, it is less affected by MUI due to the disparity in the user power levels. Hence, optimizing the the RZ thresholds provides smaller gain than for User 2.

Fig. \ref{fig:U2_SNR} shows the SER of User 2 using the proposed method compared to the benchmark schemes for different SNR conditions for the 4-QAM case. It can be seen from Fig. \ref{fig:U2_SNR} that the proposed method significantly outperforms the benchmark schemes.  The proposed method with optimized RZ thresholds provides performance gains of many orders of magnitude over the MMSE-SIC scheme and also over the naive threshold design benchmark scheme. This is because the naive threshold design with tight starting thresholds means that fewer User 1 symbols are detected during early iterations and their MUI is still present in the system when the User 2 symbols are being detected. Optimizing the thresholds to minimize User 2 MSE allows for more MUI to be removed at early iterations and improves the accuracy of User 2 symbol detection. Additionally, the proposed method provides significant performance gains over MMSE-SIC which performs very poorly, especially at high SNRs. This is due to the fact that, as the SNR increases, MMSE equalization becomes closer to zero-forcing equalization and the interference is amplified by the inverse matrix involved in the equalization process.

\begin{figure}[t]
    \centering
    \includegraphics[width = \columnwidth]{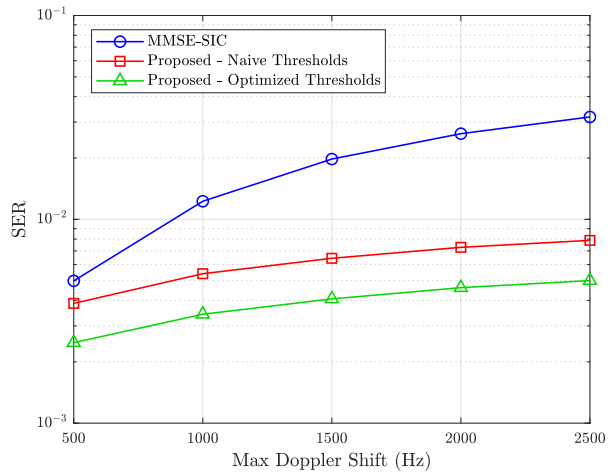}
    \caption{Comparison of the SER performance of User 1 using Algorithm 1 with optimized thresholds, Algorithm 1 with naive thresholds, and MMSE equalization with SIC, with different maximum Doppler shifts, for the case where each user is allocated a 16-QAM constellation.}
    \label{fig:U1_dopp_16}
\end{figure}

\begin{figure}[t]
    \centering
    \includegraphics[width = \columnwidth]{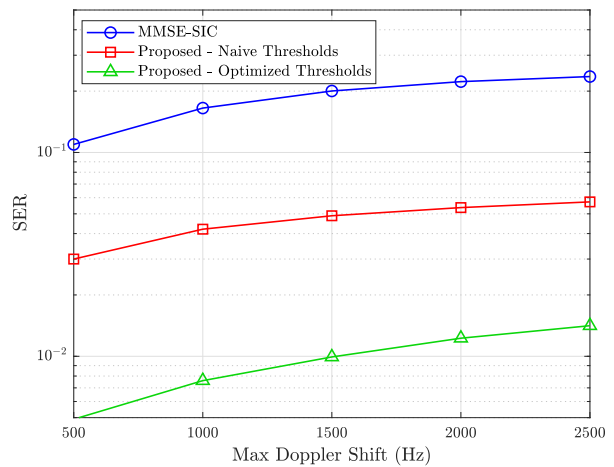}
    \caption{Comparison of the SER performance of User 2 using Algorithm 1 with optimized thresholds, Algorithm 1 with naive thresholds, and MMSE equalization with SIC, with different maximum Doppler shifts, for the case where each user is allocated a 16-QAM constellation.}
    \label{fig:U2_dopp_16}
\end{figure}

Fig. \ref{fig:U1_SNR_16} shows the SER of User 1 using the proposed equalization and detection method compared to the benchmark schemes for different SNR conditions, for the case where each user's symbols are taken from a 16-QAM constellation ($A_1 = A_2 = 16$). For these simulations, the user velocity is fixed at 200 km/h, which equates to a maximum Doppler shift of approximately 1000~Hz. It can be seen from Fig. \ref{fig:U1_SNR} that for User 1, the proposed method outperforms the MMSE-SIC method, providing an SNR gain of 6~dB. Additionally, optimizing the RZ detector thresholds provides an SNR gain of 2~dB at an SER of $10^{-2}$ over the naive threshold design benchmark. Fig. \ref{fig:U2_SNR_16} shows the SER of User 2 in OTFS-NOMA using the proposed method compared to the benchmark schemes for different SNR conditions for the 16-QAM case. It can be seen that the proposed method outperforms the benchmark the MMSE-SIC scheme for User 2 in the 16-QAM case as well. In addition, optimizing the RZ detector thresholds provides a significant performance increase over the naive threshold design benchmark scheme.


Fig. \ref{fig:U1_dopp_16} and Fig. \ref{fig:U2_dopp_16} show the SER of User 1 and User 2, respectively, under the proposed equalization and detection method, compared to the benchmark MMSE-SIC scheme, for different values of maximum Doppler shift, for the 16-QAM case. It can be seen that the performance gains of the proposed method over MMSE-SIC actually improves in high Doppler environments, as the performance of MMSE-SIC deteriorates significantly at higher maximum Doppler shifts. This is because as the Doppler shift increases, the channel matrix is more likely to be ill-conditioned; hence, the matrix inversion involved in MMSE equalization may not be robust and can introduce significant equalization error. Additionally, the optimized RZ threshold design  offers a significant performance improvement over the naive threshold design for both users. This confirms the benefits of optimizing the RZ thresholds, especially for the user with the lower power allocation.

\section{Conclusion}
This paper has presented a novel receiver for downlink OTFS-NOMA.
The proposed method uses an iterative process which deploys the LSQR algorithm to equalize the channel, RZ detection to detect symbols from both users within each iteration, and interference cancellation to remove MUI as well as IDI and ISI. The proposed modifications to the LSQR algorithm calculates the post-equalization MSE information needed for optimizing the RZ thresholds. An exact method was presented for computing the MSE as well as a low-complexity approach which takes advantage of the properties of the delay-Doppler channel in OTFS. By optimizing the thresholds, we are able to remove more MUI from the system at early iterations and are therefore able to improve detection performance on subsequent iterations. Numerical results demonstrate the superiority of the proposed method, in terms of SER performance, with respect to an MMSE-SIC benchmark scheme and with respect to a corresponding scheme with naive, pre-determined RZ threshold design.

\bibliographystyle{IEEEtran}
\bibliography{biblio}

\begin{thebibliography}{10}
\providecommand{\url}[1]{#1}
\csname url@samestyle\endcsname
\providecommand{\newblock}{\relax}
\providecommand{\bibinfo}[2]{#2}
\providecommand{\BIBentrySTDinterwordspacing}{\spaceskip=0pt\relax}
\providecommand{\BIBentryALTinterwordstretchfactor}{4}
\providecommand{\BIBentryALTinterwordspacing}{\spaceskip=\fontdimen2\font plus
\BIBentryALTinterwordstretchfactor\fontdimen3\font minus
  \fontdimen4\font\relax}
\providecommand{\BIBforeignlanguage}[2]{{%
\expandafter\ifx\csname l@#1\endcsname\relax
\typeout{** WARNING: IEEEtran.bst: No hyphenation pattern has been}%
\typeout{** loaded for the language `#1'. Using the pattern for}%
\typeout{** the default language instead.}%
\else
\language=\csname l@#1\endcsname
\fi
#2}}
\providecommand{\BIBdecl}{\relax}
\BIBdecl

\bibitem{Tataria_6G}
H.~Tataria, M.~Shafi, A.~F. Molisch, M.~Dohler, H.~Sjöland, and F.~Tufvesson,
  ``{6G Wireless Systems: Vision, Requirements, Challenges, Insights, and
  Opportunities},'' \emph{Proceedings of the IEEE}, vol. 109, no.~7, pp.
  1166--1199, 2021.

\bibitem{Wei_OTFS}
Z.~Wei, W.~Yuan, S.~Li, J.~Yuan, G.~Bharatula, R.~Hadani, and L.~Hanzo,
  ``{Orthogonal Time-Frequency Space Modulation: A Promising Next-Generation
  Waveform},'' \emph{IEEE Wireless Communications}, vol.~28, no.~4, pp.
  136--144, 2021.

\bibitem{Hadani_OTFS}
R.~Hadani, S.~Rakib, M.~Tsatsanis, A.~Monk, A.~J. Goldsmith, A.~F. Molisch, and
  R.~Calderbank, ``{Orthogonal Time Frequency Space Modulation},'' in
  \emph{{2017 IEEE Wireless Communications and Networking Conference (WCNC)}},
  2017, pp. 1--6.

\bibitem{Tiwar_OTFS_LMMSE}
S.~Tiwari, S.~S. Das, and V.~Rangamgari, ``{Low complexity LMMSE Receiver for
  OTFS},'' \emph{IEEE Communications Letters}, vol.~23, no.~12, pp. 2205--2209,
  2019.

\bibitem{Surabhi_OTFS_EQ}
G.~D. Surabhi and A.~Chockalingam, ``{Low-Complexity Linear Equalization for
  OTFS Modulation},'' \emph{IEEE Communications Letters}, vol.~24, no.~2, pp.
  330--334, 2020.

\bibitem{ZOU_OTFS_EQ}
T.~Zou, W.~Xu, H.~Gao, Z.~Bie, Z.~Feng, and Z.~Ding, ``{Low-Complexity Linear
  Equalization for OTFS Systems with Rectangular Waveforms},'' in \emph{IEEE
  International Conference on Communications Workshops}, 2021, pp. 1--6.

\bibitem{Rav_INT_canc_MP}
P.~Raviteja, K.~T. Phan, Y.~Hong, and E.~Viterbo, ``{Interference Cancellation
  and Iterative Detection for Orthogonal Time Frequency Space Modulation},''
  \emph{IEEE Transactions on Wireless Communications}, vol.~17, no.~10, pp.
  6501--6515, 2018.

\bibitem{Surabhi_MP_mmWAVE}
G.~D. Surabhi, M.~K. Ramachandran, and A.~Chockalingam, ``{OTFS Modulation with
  Phase Noise in mmWave Communications},'' in \emph{2019 IEEE 89th Vehicular
  Technology Conference (VTC2019-Spring)}, 2019, pp. 1--5.

\bibitem{MIMO_OTFS_MP}
M.~Kollengode~Ramachandran and A.~Chockalingam, ``{MIMO-OTFS in High-Doppler
  Fading Channels: Signal Detection and Channel Estimation},'' in \emph{2018
  IEEE Global Communications Conference (GLOBECOM)}, 2018, pp. 206--212.

\bibitem{Qu_OTFS_detection}
H.~Qu, G.~Liu, L.~Zhang, S.~Wen, and M.~A. Imran, ``{Low-Complexity Symbol
  Detection and Interference Cancellation for OTFS System},'' \emph{IEEE
  Transactions on Communications}, vol.~69, no.~3, pp. 1524--1537, 2021.

\bibitem{patent_Hadani}
H.~R. Rakib~Shlomo, ``{Multiple access in wireless telecommunications system
  for high-mobility applications},'' August 2017.

\bibitem{surabhi2019multiple}
G.~D. Surabhi, R.~M. Augustine, and A.~Chockalingam, ``{Multiple Access in the
  Delay-Doppler Domain using OTFS modulation},'' 2019.

\bibitem{Chong_OTFS_Uplink_Rate}
R.~Chong, S.~Li, J.~Yuan, and D.~W.~K. Ng, ``{Achievable Rate Upper-Bounds of
  Uplink Multiuser OTFS Transmissions},'' \emph{IEEE Wireless Communications
  Letters}, vol.~11, no.~4, pp. 791--795, 2022.

\bibitem{DaiSurveyNOMA}
L.~{Dai}, B.~{Wang}, Z.~{Ding}, Z.~{Wang}, S.~{Chen}, and L.~{Hanzo}, ``{A
  Survey of Non-Orthogonal Multiple Access for 5G},'' \emph{IEEE Communications
  Surveys Tutorials}, vol.~20, no.~3, pp. 2294--2323, 2018.

\bibitem{Poor_OTFS_NOMA}
Z.~Ding, R.~Schober, P.~Fan, and H.~Vincent~Poor, ``{OTFS-NOMA: An Efficient
  Approach for Exploiting Heterogenous User Mobility Profiles},'' \emph{IEEE
  Transactions on Communications}, vol.~67, no.~11, pp. 7950--7965, 2019.

\bibitem{Chatt_OTFS_NOMA}
A.~Chatterjee, V.~Rangamgari, S.~Tiwari, and S.~S. Das, ``{Nonorthogonal
  Multiple Access With Orthogonal Time–Frequency Space Signal
  Transmission},'' \emph{IEEE Systems Journal}, vol.~15, no.~1, pp. 383--394,
  2021.

\bibitem{OTFS_SCMA}
K.~Deka, A.~Thomas, and S.~Sharma, ``{OTFS-SCMA: A Code-Domain NOMA Approach
  for Orthogonal Time Frequency Space Modulation},'' \emph{IEEE Transactions on
  Communications}, vol.~69, no.~8, pp. 5043--5058, 2021.

\bibitem{OTFS_code_noma}
H.~Wen, W.~Yuan, and S.~Li, ``{Downlink OTFS Non-Orthogonal Multiple Access
  Receiver Design based on Cross-Domain Detection},'' in \emph{IEEE
  International Conference on Communications Workshops}, 2022, pp. 928--933.

\bibitem{Taubock_LSQR_2011}
G.~Taubock, M.~Hampejs, P.~Svac, G.~Matz, F.~Hlawatsch, and K.~Grochenig,
  ``{Low-Complexity ICI/ISI Equalization in Doubly Dispersive Multicarrier
  Systems Using a Decision-Feedback LSQR Algorithm},'' \emph{IEEE Transactions
  on Signal Processing}, vol.~59, no.~5, pp. 2432--2436, 2011.

\bibitem{mcwade_OTFS}
\BIBentryALTinterwordspacing
S.~McWade, M.~F. Flanagan, and A.~Farhang, ``{Low-Complexity Equalization and
  Detection for OTFS-NOMA},'' 2022. [Online]. Available:
  \url{https://arxiv.org/abs/2211.07388}
\BIBentrySTDinterwordspacing

\bibitem{Arman_OTFS_2018}
A.~Farhang, A.~RezazadehReyhani, L.~E. Doyle, and B.~Farhang-Boroujeny, ``{Low
  Complexity Modem Structure for OFDM-Based Orthogonal Time Frequency Space
  Modulation},'' \emph{IEEE Wireless Communications Letters}, vol.~7, no.~3,
  pp. 344--347, 2018.

\bibitem{Hampeis_LSQR}
M.~Hampejs, P.~Svac, G.~Taubock, K.~Grochenig, F.~Hlawatsch, and G.~Matz,
  ``{Sequential LSQR-based ICI equalization and decision-feedback ISI
  cancellation in pulse-shaped multicarrier systems},'' in \emph{IEEE 10th
  Workshop on Signal Processing Advances in Wireless Communications}, 2009, pp.
  1--5.

\bibitem{LSQR}
C.~C. Paige and M.~A. Saunders, ``{LSQR: An Algorithm for Sparse Linear
  Equations and Sparse Least Squares},'' \emph{ACM Trans. Math. Softw.},
  vol.~8, no.~1, p. 43–71, mar 1982.

\bibitem{Hrycak_LSQR_GMRES}
T.~Hrycak, S.~Das, G.~Matz, and H.~G. Feichtinger, ``{Low Complexity
  Equalization for Doubly Selective Channels Modeled by a Basis Expansion},''
  \emph{IEEE Transactions on Signal Processing}, vol.~58, no.~11, pp.
  5706--5719, 2010.

\bibitem{QU_ScFDMA}
H.~Qu, G.~Liu, Y.~Wang, Q.~Chen, C.~Yi, and J.~Peng, ``{A Time-Domain Approach
  to Channel Estimation and Equalization for the SC-FDM System},'' \emph{IEEE
  Transactions on Broadcasting}, vol.~65, no.~4, pp. 713--726, 2019.

\bibitem{Yin_CG_mod}
B.~Yin, M.~Wu, J.~R. Cavallaro, and C.~Studer, ``{Conjugate gradient-based
  soft-output detection and precoding in massive MIMO systems},'' in \emph{IEEE
  Global Communications Conference}, 2014, pp. 3696--3701.

\bibitem{He_NOMA_SER_2019}
Q.~He, Y.~Hu, and A.~Schmeink, ``{Closed-Form Symbol Error Rate Expressions for
  Non-Orthogonal Multiple Access Systems},'' \emph{IEEE Transactions on
  Vehicular Technology}, vol.~68, no.~7, pp. 6775--6789, 2019.

\bibitem{brent71:CJ-14-422}
R.~P. Brent, ``{An Algorithm with Guaranteed Convergence for Finding a Zero of
  a Function},'' \emph{The Computer Journal}, vol.~14, pp. 422--425, 1971.

\bibitem{3gpp_TS38901}
\emph{3GPP TS 38.901}, 3rd Generation Partnership Project (3GPP), June 2018,
  v15.0.0.

\bibitem{jakes_model}
C.~Xiao, Y.~Zheng, and N.~Beaulieu, ``{Second-order statistical properties of
  the WSS Jakes’ fading channe simulator},'' \emph{IEEE Transactions on
  Communications}, vol.~50, pp. 888 -- 891, 07 2002.

\end{thebibliography}

\end{document}